\documentclass[longbibliography,floatfix,aps,reprint,superscriptaddress,amsmath,amssymb,prx,footinbib]{revtex4-2}
\usepackage{amsmath}
\usepackage{graphicx}
\usepackage{dcolumn}
\usepackage{bm}
\usepackage{float}
\usepackage[english]{babel} 
\usepackage[usenames,dvipsnames]{xcolor}
\usepackage[colorlinks=true,citecolor=Cerulean,linkcolor=RubineRed,urlcolor=Cerulean]{hyperref}
\usepackage{booktabs}
\usepackage{mathtools}
\usepackage{physics}

\newcommand{\betabb}{\boldsymbol{\beta}}

\begin{document}

\title{Counterdiabatic Optimised Local Driving}
\author{Ieva \v{C}epait\.{e}}
\email{ieva.cepaite@strath.ac.uk}
\affiliation{Department of Physics, SUPA and University of Strathclyde, Glasgow G4 0NG, United Kingdom}
\author{Anatoli Polkovnikov}
\affiliation{Department of Physics, Boston University, Boston, Massachusetts 02215, USA}
\author{Andrew J. Daley}
\affiliation{Department of Physics, SUPA and University of Strathclyde, Glasgow G4 0NG, United Kingdom}
\author{Callum W. Duncan}
\affiliation{Department of Physics, SUPA and University of Strathclyde, Glasgow G4 0NG, United Kingdom}

\date{\today}

\begin{abstract}

Adiabatic protocols are employed across a variety of quantum technologies, from implementing state preparation and individual operations that are building blocks of larger devices, to higher-level protocols in quantum annealing and adiabatic quantum computation. The problem of speeding up these processes has garnered a large amount of interest,  resulting in a menagerie of approaches, most notably quantum optimal control and shortcuts to adiabaticity. The two approaches are complementary: optimal control manipulates control fields to steer the dynamics in the minimum allowed time while shortcuts to adiabaticity aim to retain the adiabatic condition upon speed-up. We outline a new method which combines the two methodologies and takes advantage of the strengths of each. The new technique improves upon approximate local counterdiabatic driving with the addition of time-dependent control fields.  We refer to this new method as counterdiabatic optimised local driving (COLD) and we show that it can result in a substantial improvement when applied to annealing protocols,  state preparation schemes, entanglement generation and population transfer on a lattice. We also demonstrate a new approach to the optimisation of control fields which does not require access to the wavefunction or the computation of system dynamics. COLD can be enhanced with existing advanced optimal control methods and we explore this using the chopped randomised basis method and gradient ascent pulse engineering. 

\end{abstract}

\maketitle

\section{Introduction}\label{sec:intro}

Time-dependent manipulation of few and many-particle quantum systems is important across all implementations of quantum computing and simulation. In such processes, decoherence and undesired transitions reducing the state fidelity are relatively ubiquitous. One important example is given by the undesired transitions that can occur between instantaneous eigenstates of the dynamical Hamiltonian upon the application of an external drive. This is why many driving protocols rely on adiabatic dynamics, where the system follows the instantaneous eigenstates and transitions are naturally suppressed.  Ideal adiabatic processes are reversible making them - in principle - robust. However, to approach ideal adiabatic processes the dynamics must always be very slow, requiring compromises on the time-scales of competing heating and decoherence processes. 

Speeding up adiabatic protocols to enable their completion within the system's coherence time is important for the development of any quantum technologies relying on such protocols \cite{acin2018quantum}.  One approach to do this is the implementation of optimal driving protocols, which aim to end up with the system in a desired final state.  For example, numerically optimised paths can be employed to avoid points where gaps in the spectrum of the system become small, or additional control fields can be tuned to increase the size of these gaps \cite{kirk2004optimal,glaser2015training,AlessandroBook2007}. In broad terms, this is the goal of protocols  collectively referred to as quantum optimal control. Another option is to design techniques which speed up the adiabatic dynamics, often termed shortcuts to adiabaticity (STA). The primary aim of STA is to entirely remove or suppress diabatic transitions between instantaneous eigenstates of the dynamical Hamiltonian \cite{Torrontegui2013,GueryOdelin2019}. One particularly successful technique is counterdiabatic driving (CD), which was first utilised in physical chemistry by Demirplak and Rice \cite{demirplak2003,demirplak2005}, and was independently introduced by Berry \cite{berry_transitionless_2009} under the name `transitionless driving'. CD aims to suppress losses that arise due to fast deformations of the system far from the adiabatic limit by analytically compensating for them in the Hamiltonian.  In general,  to suppress diabatic losses exactly,  the full analytical or numerical solutions of the Schr\"odinger equation are required. This makes the implementation of CD in complex systems - \@e.g.~for many-body dynamics - difficult and requires the need for new techniques to be introduced.

Links between optimal control and STA have existed throughout the development of both approaches  \cite{stefanatos2010frictionless, stefanatos2021,Zhang2021connection},  but there are few examples of their explicit combination in a way that exploits their complementary nature.  Some attempts to achieve this have included an emulation of CD through fast oscillations of the original Hamiltonian \cite{Petiziol2018fast, Petiziol2019accelerating} as well as through recent advances in reinforcement learning methods aimed at optimizing quantum protocols~\cite{bukov_reinforcement_2018}.  Such methods have been shown to achieve a significant improvement in performance when implemented using concepts borrowed from CD~\cite{yao_reinforcement_2020}. In this work, we offer a significantly different new approach in combining elements from STA and quantum optimal control which we will call \textit{counterdiabatic optimised local driving} (COLD). 

A key ingredient in the development of COLD is a recent approach designed for implementing CD in the setting of larger, more complex systems: local counterdiabatic driving (LCD) \cite{sels_minimizing_2017,Kolodrubetz2017geometry,  gjonbalaj2021counterdiabatic}.  LCD offers a method to derive \emph{approximate} CD protocols, with the aim of suppressing undesired transitions instead of fully eliminating them. This allows it to account for some physical constraints of the system,  \@e.g.~locality conditions.  However, the approximate nature of the LCD protocol can lead to poor performance,  necessitating the introduction of additional non-local, long-range corrections \cite{Kyaw2018cluster, delcampo2012assisted, sels_minimizing_2017}. If all possible corrections are added, then LCD is equivalent to the normal analytical approaches of CD, but the additional terms are generally difficult to control experimentally.  COLD offers an alternative approach,  with additional control fields which allow for an optimisation of the dynamical Hamiltonian for a given local form of LCD. The impact of more complex corrections can then be radically reduced,  giving a corresponding improvement in the desired protocol. 

An important consequence of optimising for a given local order of the LCD is the possibility of bypassing the need to have access to the wave function, dynamics or experimental data of the given system in order to perform the optimisation. LCD is an analytic method and can be calculated using only the coefficients of the Hamiltonian. We find that it is possible to perform numerical optimisation of the path of the system by simply minimising higher order LCD integrals and/or amplitudes, a method that not only bypasses the need to compute system dynamics but is also independent of system size. This makes it an exceptionally useful tool in practical settings.

The structure of this paper is as follows: first, we give a detailed description of the new method, COLD, with a focus on the elements of quantum optimal control and CD required for its implementation.  In Sec.~\ref{sec:TwoSpin} we explore a 2-spin annealing protocol,  that showcases the strengths of COLD.  Sec.~\ref{sec:1dIsing} describes and analyses the improvements gained with COLD and its combination with other optimal control techniques in the case of state preparation in the Ising model as well as the potential computational advantage of optimising higher order integrals of LCD instead of the final state fidelity.  In Sec.~\ref{sec:lattice} we show the improvement that COLD can achieve on the recently realised example of LCD for state transfer on a synthetic lattice in ultracold atoms and in Sec.~\ref{sec:ghz} we demonstrate that, when implemented for second order LCD, COLD can be used to quickly and effectively prepare highly entangled multipartite GHZ states.  Finally, in Sec.~\ref{sec:minimisation} we explore the possibility of minimising both the power and amplitude of higher order LCD drives as a means to efficiently optimise COLD parameters, bypassing the requirement of computing system dynamics. A list of abbreviations used in this work can be found in Table.~\ref{table} for reference. 

\begin{table}[h]
      \begin{tabular}{p{2cm} | p{6cm}} \toprule
         \multicolumn{1}{m{2cm}}{\centering Abbreviation}
         & \multicolumn{1}{m{6cm}}{\centering Meaning} \\
         \midrule
         STA & shortcuts to adiabaticity \\ \hline
         CD & counterdiabatic driving \\ \hline
         LCD & local counterdiabatic driving \\ \hline
         COLD & counterdiabatic optimised local driving \\ \hline
         BPO & bare Powell optimisation \\ \hline
         BDA & bare dual annealing \\ \hline
         CRAB & chopped randomised basis \\ \hline
         GRAPE & gradient ascent pulse engineering \\ \hline
         ARP & adiabatic rapid passage \\
         \bottomrule
      \end{tabular}
      \caption{List of abbreviations used throughout the manuscript.}
\end{table}\label{table}

\section{An Introduction to Counterdiabatic Optimised Local Driving}\label{sec:intro_olcd}

\subsection{Quantum Optimal Control}\label{sec:OptCont}

 In the context we consider,  we employ quantum optimal control to optimise the function $f(\psi,\betabb)$ in the Schr\"odinger equation
\begin{equation}
\dot{\psi} = f(\psi,\betabb),
\label{eq:Control}
\end{equation}
where $\psi$ is the quantum wave function and $\betabb$ is the set of optimisable control parameters. Optimisation of Eq.~\eqref{eq:Control} in most cases means taking the system from an initial state $\ket{\psi_0}$ to a final target state $\ket{\psi_T}$ by finding the optimal values of $\betabb$ with respect to some target metric (e.g.~the time taken to evolve the system from $\ket{\psi_0}$  to $\ket{\psi_T}$).  There is a large variety of techniques available to achieve this goal \cite{glaser2015training,koch2016controlling}. %In all numerical optimisations, including for the COLD method, we will use the Powell minimisation approach \cite{powell1964efficient} unless stated otherwise.  

The success/target metric needs to be defined prior to the optimisation of $\betabb$. Often this is done by constructing a \emph{cost function}, which in turn defines the optimisation landscape.  In general,  we can optimise for any desired property of the final state of the system, with some examples being the entropy,  energy,   energy fluctuations or some other observable.  A commonly used cost function in state preparation is related  to the fidelity of the final, post-evolution state $\ket{\psi_f(\boldsymbol \beta)}$ with respect to the target state:
\begin{align} \label{eq:lossfunc}
    \mathcal{C}(\boldsymbol \beta) = 1 - \left|\braket{\psi_T}{\psi_f(\boldsymbol \beta)}\right|^2.
\end{align}
In performing such a numerical optimisation,  it is common to take the target state to be parameterised via a Hamiltonian split into two parts. The first is the so-called \emph{bare} Hamiltonian $H_0(t)$, which can be time-dependent and describes the dynamics of the quantum system in question. The second part is then an additional driving term that includes a function $f$ parameterised by the control parameters $\betabb$, as well as operators $\mathcal{O}_{\rm opt}$ which provide additional degrees of freedom in the dynamics of the system. The full Hamiltonian of the control system is then:
\begin{align} \label{eq:h_optimal_control}
    H_{\beta}(t,\betabb)  = H_0(t) + f(t, \betabb)\mathcal{O}_{\rm opt}.
\end{align}
The parameters $\betabb$ can then be optimised for the optimal dynamics with respect to the metric defined by the cost function.

In this work,  we generally use the Powell minimization  \cite{powell1964efficient} and dual annealing \cite{XIANG1997generalized} approaches for the numerical optimisation as implemented in Python's \textit{SciPy} \cite{Pauli2020SciPy}.  When performing this optimisation without any CD terms in the Hamiltonian, we refer to them as bare Powell optimisation (BPO) and bare dual annealing (BDA) respectively, with bare referring to the lack of CD. %This is the simplest and least resource-intensive approach in general,  since it requires neither the additional drives of the LCD nor the potential computational intensiveness of quantum optimal control approaches. This makes BPO a good comparison case for COLD, as the fidelities reached for state preparation need to be substantially improved beyond those achieved by BPO to warrant the engineering required for COLD.
Furthermore, we implement the chopped randomised basis (CRAB) approach \cite{Caneva_chopped_2011,muller2021} and combine its methodology with that of COLD, to obtain the method of COLD-CRAB.  CRAB expands the size of the parameter landscape by employing randomisation,  usually in the optimised pulse driving the system.  The approach was first developed for quantum many-body problems whose simulation requires the time-dependent density matrix renormalization group, despite the fact that these were thought to not be tractable in the quantum control setting \cite{brif2010,muller2021}.  CRAB has benefits in that it can avoid traps in the control landscape \cite{Rach2015}, and has built-in flexibility for open-loop or closed-loop optimisation \cite{heck2018,muller2021} although these advantages come at a higher computational cost due to requiring a far larger search-space for the optimisation. 

\subsection{Counterdiabatic Driving}\label{sec:LCD}

An important class of optimisation problems deals with the case where the initial and final states are ground states of a Hamiltonian $H_0(t)$ at some initial $t=t_i$ and final $t=t_f$ time.  In these cases, the adiabatic theorem guarantees that for an infinitesimally slow transformation of the system $t_f-t_i\to\infty$,  it should follow the instantaneous (non-degenerate) ground state of $H_0(t)$ and hence reach the target state with unit fidelity. This process is generally known as quantum annealing. 

In large,  complex systems with many degrees of freedom, quantum annealing tends to require prohibitively long protocol times due to vanishingly small gaps typically present in such systems. This often makes annealing protocols impractical \cite{farhi2008how, Wurtz2022counterdiabaticity}.  It has been found that this problem can be formally overcome by using CD,  where velocity-dependent terms are added to the Hamiltonian analytically enforcing the adiabatic wave function to be the solution of the time-dependent Schr\"odinger equation \cite{demirplak2003,demirplak2005,berry_transitionless_2009}.  In this case, the dynamical state will follow the instantaneous eigenstate with no transitions regardless of the driving time.  The form of the dynamical Hamiltonian enforcing this is \cite{berry_transitionless_2009}:
\begin{equation}\label{eq:counterdiabatic}
\begin{aligned}
    & H_{\mathrm{CD}}(t) = H_0 (t) \\ & + i\hbar \sum_n (\ket{\partial_t n}\bra{n} - \bra{n}\ket{\partial_t n}\ket{n}\bra{n}),
\end{aligned}
\end{equation}
with $\ket{n}\equiv \ket{n(t)}$ the $n$-th eigenstate of the instantaneous Hamiltonian $H_0(t)$. The last term enforces the phases ($\bra{n}\ket{\partial_t n}$) on the instantaneous eigenstates, which are arbitrary and thus will be omitted.  In general,  knowledge of the CD Hamiltonian of Eq.~\eqref{eq:counterdiabatic} requires knowledge of the full spectrum of $H_0(t)$ at each instantaneous moment in time.

\subsection{Counterdiabatic Optimised Local Driving} \label{sec:OLCD}

We will now introduce the main idea of COLD. The principle is to take the same approach as Sec.~\ref{sec:LCD} but with the original Hamiltonian given by $H_\beta(t,\betabb)$, see Eq.~\eqref{eq:h_optimal_control}. Quantum Annealing then applies to the whole family of control Hamiltonians $ H_{\beta}(t,\betabb)$ as long as the additional control function $f(t, \betabb)$ vanishes at the protocol boundaries: $f(t_i, \betabb)=f(t_f, \betabb)=0$.  This flexibility was explored in finding the optimal adiabatic path characterized by e.g. the shortest distance between the initial and the final states, i.e. a geodesic \cite{tomka2016geodesic}. A similar geodesic approach was developed in the context of dissipative systems to minimize energy losses~\cite{Sivak2012Geodesic}.  During the protocol,  a dynamical Hamiltonian $H_\beta(t,\betabb)$ generally induces transitions between the quantum states that it drives and the question about what is the optimal path remains open.  

The Hamiltonian $H_\beta(t,\betabb)$ and its eigenstates depend on time only through the driving parameters, which include $\betabb$ and any additional control terms in the particular protocol. This makes it convenient to introduce a path in the coupling space parametrized by a dimensionless parameter $\lambda\in [0,1]$ such that both $H_0$ and $f$ are functions of $\lambda$ satisfying $H_\beta(\lambda=0)=H_{0}(t_i)$ and $H_\beta(\lambda=1)=H_{0}(t_f)$,  i.e.  being equal to the initial and the final Hamiltonian at the protocol boundaries.  By construction this implies that any additional fields introduced to the bare Hamiltonian $H_0$ must go to zero at the boundaries.  In this way,  any protocol can be uniquely characterized by first specifying the path $f(\lambda, \betabb)$ in the coupling space manifold and then the time dependence $\lambda(t)$ along it. The path determines the sequence of couplings of the Hamiltonian during time evolution and hence the sequence of ground state wave functions followed by the driven state.  Furthermore, the time dependence encodes the speed of traversing this path.  We can then introduce a hermitian operator called the (path-dependent) adiabatic gauge potential~\cite{sels_minimizing_2017}: $\mathcal A_\lambda=i \hbar \sum_n \ket{\partial_\lambda  n}\bra{n}$,  which satisfies a closed form equation,
\begin{equation}
\label{eq:AGP_eq}
[ G_{\lambda},H_\beta ] = 0,
\end{equation}
where:
\begin{equation}\label{eq:Goperator}
G_{\lambda} = \partial_\lambda H_\beta+{i\over \hbar} [ \mathcal A_\lambda,H_\beta],
\end{equation}
with both $H_{\beta}$ and $\mathcal{A}$ having a dependence on $\lambda$ and $\betabb(\lambda)$. We note that in the case of a nonlinear Schr\"odinger equation where the dynamics are described by classical Hamiltonian equations of motion, the commutators need only be replaced with Poisson brackets and the same idea applies \cite{gjonbalaj2021counterdiabatic}.

Thus, the CD Hamiltonian reads
\begin{align}\label{eq:counterdiabaticLCD}
    H_{\mathrm{CD}}(\lambda,\betabb) = H_\beta(\lambda,\betabb) +\dot \lambda \mathcal A_\lambda(\lambda,\betabb),
\end{align}
and is equivalent to the CD Hamiltonian of Eq.~\eqref{eq:counterdiabatic} given knowledge of the exact adiabatic gauge potential. However, generally the adiabatic gauge potential is a very non-local object and solutions of Eq.~\eqref{eq:AGP_eq} are unstable to small perturbations containing exponentially many terms in the number of degrees of freedom. 

LCD aims to find approximate gauge potentials that satisfy particular requirements like robustness and locality,  thus circumventing many of the difficulties in determining the second component in Eq.~\eqref{eq:counterdiabatic} and~\eqref{eq:counterdiabaticLCD} exactly. The goal, in essence, is to suppress the most relevant diabatic effects rather than completely eliminate them.  This method has recently been experimentally implemented to speed up state transfer for synthetic lattices in ultracold atoms \cite{meier_counterdiabatic_2020}, for preparing states in nuclear-magnetic-resonance systems \cite{zhou_experimental_2020}, and annealing protocols on an IBM quantum computer \cite{Hegade2021Shortcuts, Wurtz2022counterdiabaticity}.

Following the methods of Ref.~\cite{sels_minimizing_2017}, the problem of finding the optimal adiabatic gauge potential can be cast as the minimisation of the Hilbert-Schmidt norm of $G_\lambda$, which is equivalent to minimisation of the action
\begin{equation}\label{eq:actionCD}
\mathcal{S}(\mathcal{A}_{\lambda}) = \Trace{\left[G_{\lambda}(\mathcal{A}_{\lambda})^2\right]},
\end{equation}
with respect to $\mathcal{A}_{\lambda}$. In most cases, this is achieved by first choosing an operator ansatz - \@i.e. a set of linearly independent operators $\{\mathcal{O}_{\rm LCD}\}$ - and then using this set as an operator basis for the adiabatic gauge potential $\mathcal A_\lambda=\sum_j \alpha_j \mathcal O_{\rm LCD}^{(j)}$. The action can then be minimized with respect to the the set of coefficients, ${\bm \alpha}$.  The choice of operators $\{\mathcal{O}_{\rm LCD}\}$ can be made easier when noting that $\mathcal{A}_{\lambda}$ acts as a generators of motion in the parameter space. This implies that for, say, real Hamiltonians like those we'll be exploring in the following sections, a good Ansatz for the adiabatic gauge potential is one which is non-interacting and \emph{imaginary}. In the example of an 
Ising spin chain we may take $\mathcal{A}_\lambda = \sum_j^N \alpha_j \sigma^y_j$, where $j$ labels the $N$ chain sites, and $\{\mathcal{O}_{\rm LCD}\}$ is a set the $y$-pauli matrices.  

Without any additional control fields $f(\lambda,\betabb)$, LCD is essentially an informed choice of the operator set $\{\mathcal{O}_{\rm LCD}\}$ in a way that the resulting control protocol from the minimisation of Eq.~\eqref{eq:actionCD} is optimal for a given $H_0(\lambda)$. In this case we explore the family of Hamiltonians
\begin{equation}
H_{\rm LCD}(\lambda) = H_0(\lambda) +\sum_j \alpha_j(\lambda) \mathcal{O}_{\rm LCD}^{(j)}.
\end{equation}
The performance of such LCD protocols is determined by how accurately the variational manifold spanned by the set $\{\mathcal{O}_{\rm LCD}\}$ can approximate an exact $\mathcal{A}_{\lambda}$ such that Eq.~\eqref{eq:AGP_eq} holds. 

In the case of the new protocol COLD,  we allow for extra exploration of the family of Hamiltonians due to the additional control fields as in Eq.~\eqref{eq:h_optimal_control}.  This expands the family of Hamiltonians to
\begin{equation}\label{eq:expandedH}
\begin{split}
H_{\rm COLD}(\lambda,\betabb) &= H_0(\lambda) + {\bm \alpha}(\lambda,\betabb) \mathcal{O}_{\rm LCD} \\
&+ f(\lambda,\betabb) \mathcal{O}_{\rm opt}.
\end{split}
\end{equation}
Note, that the coefficients of the optimal control field change the form of the LCD driving coefficients, i.e. $\bm \alpha \rightarrow \alpha(\lambda,\betabb)$.  The aim of COLD is then to optimise the coefficients $\betabb$ in such a way that the LCD term in the above equation allows for the greatest suppression of non-adiabatic effects for the dynamical Hamiltonian $H_0(\lambda)+f(\lambda,\betabb)\mathcal{O}_{\rm opt}$. One can picture it as changing the path that the system takes between its initial and final states, with the express goal of picking a path that maximises the effects of the approximate counterdiabatic drive given by the second term in the equation. This path will depend on the form of the optimal pulse function, the operators $\mathcal{O}_{\rm opt}$ and on the values of $\betabb$. We will focus on the optimisation of the control parameters $\betabb$ for a given choice of $f(\lambda, \betabb)$ and $\mathcal{O}_{\rm opt}$, although the choice of operators $\mathcal{O}_{\rm opt}$ as well as the function of the control pulse $f(\lambda, \betabb)$ can also be optimised over as an extension.

With COLD, we have two methods to improve on the existing LCD protocol.  As previously shown in Refs.~\cite{claeys_floquet-engineering_2019,prielinger_diabatic_2020}, there is a possibility to add more terms to the LCD making it less local, \@e.g.~through long-range interactions.  In the spin chain case, we could take the aforementioned sum over $\sigma^y$ terms to be the \textit{first-order} anzatz for the LCD, where higher-order ans\"{a}tze might contain sets of operators $\{\mathcal{O}_{\rm LCD}\}$ with terms odd in $\sigma^y$ such as $\sigma^y_j\sigma^{(z,x)}_{j+1}$. This procedure generally improves the performance of CD protocols at the expense of adding more complex operators which may be experimentally impractical depending on the scenario.  Alternately, with COLD and the introduction of additional local control fields to the Hamiltonian,  we can improve the performance of LCD at a fixed complexity of the CD term by significantly modifying the adiabatic landscape at intermediate couplings to enhance the performance of the given order of LCD. 

In this work we pursue two directions of optimising the local control fields: numerical optimisation of the dynamics and minimisation of the higher order LCD terms. For the most part we will focus on numerical optimisation of the dynamics directly, as these will reach optimal values for specific protocols when implemented. COLD opens the possibility of minimising the higher order LCD terms instead, which benefits from not requiring calculation of the systems dynamics. This approach, as discussed in Sec.~\ref{sec:minimisation}, allows for optimal control procedures using COLD to be implemented for large systems that would be cumbersome for procedures that require the numerical optimisation of the dynamics.

We also note that while it may seem prudent to treat the LCD coefficients $\alpha(t)$ as control pulses which may be parameterised and optimised in the same way that $f$ is, we find that this method fails to perform well compared to using the variational form of the LCD as we have done. This is likely due to a difficulty in choosing an accurate form of the drive as well as parameterising it. On top of that, the loss function space in this case may become intractable when compared to that of COLD as we have presented it in this section.

Furthermore, we compare COLD to the use of CRAB, as discussed in Sec.~\ref{sec:OptCont}.  An advantage of COLD is that it can be combined with many advanced optimal control procedures, owing to the standard way additional control fields are introduced to the Hamiltonian. In this work we find the combination of COLD and CRAB particularly useful and we will refer to this as COLD-CRAB.

\section{Two Spin Quantum Annealing}\label{sec:TwoSpin}

To showcase and explore the use of COLD in a relatively simple setting we will consider a two spin quantum annealing problem with bare Hamiltonian
\begin{equation}\label{eq:Hanneal}
H_0(t) = -2J \sigma_1^z \sigma_2^z - h ( \sigma_{1}^z + \sigma_{2}^z) +  2h \lambda(t) (\sigma_{1}^x + \sigma_{2}^x),
\end{equation}
where $\sigma^a_j$, $a \in \{x,y,z\}$ are the Pauli matrices applied to spins indexed by $j$.  For the scaling function $\lambda(t)$ we pick the form 
\begin{equation}\label{eq:Scalingfunc}
\lambda(t) = \sin^2\left(\frac{\pi}{2} \sin^2 \left( \frac{\pi t}{2 \tau} \right) \right)
\end{equation}
such that $\lambda(0) = 0$ and $\lambda(\tau) = 1$. We consider the case of $J/h=0.5$,  which means the spins start in the initial state of $\ket{\uparrow \uparrow}$ and finish in a superposition of all of the symmetric states.  

As discussed in Ref.~\citep{sels_minimizing_2017}, since $H_0$ has a standard Ising spin chain form, the first-order LCD terms are given by the following ansatz for the adiabatic gauge potential:
\begin{equation}\label{eq:LCD1st}
\mathcal{A}(\lambda) = \alpha \sum_{i=1}^2 \sigma_i^y,
\end{equation}
with the sum being over the full length of the $N$ spin chain.  Minimising Eq.~\eqref{eq:actionCD} for this $\mathcal{A}_{\lambda}$ with respect to the coefficient $\alpha$ gives
\begin{equation}
\alpha = - \frac{h^2}{4(h\lambda)^2 + h^2 + 4J^2}.
\end{equation}

To further improve on the first-order LCD we can implement COLD, as we will discuss shortly, or we can introduce higher-order terms to the ansatz for $\mathcal{A}_{\lambda}$. This second method serves as a good benchmark against COLD, since it offers an improvement to first-order LCD in the same way as COLD does, but requires more complicated interactions between the two spins increasing the implementation overhead.  The second-order LCD can be found by taking an ansatz for the adiabatic gauge potential:
\begin{equation}
\begin{aligned}
 \mathcal{A}^{(2)}(\lambda) =& \alpha \sum_{j} \sigma_j^y +  \gamma (\sigma_1^x \sigma_{2}^y + \sigma_1^y \sigma_{2}^x) \\ & +  \zeta (\sigma_1^z \sigma_{2}^y + \sigma_1^y \sigma_{2}^z),
 \end{aligned} 
\end{equation}
where to solve for $\alpha$, $\gamma$, and $\zeta$ we once again minimize the action given by Eq.~\eqref{eq:actionCD} and obtain three coupled equations which can be solved numerically (see Appendix \ref{app:derivation} for a detailed derivation).

\begin{figure}[t]
	\includegraphics[width=0.98\linewidth]{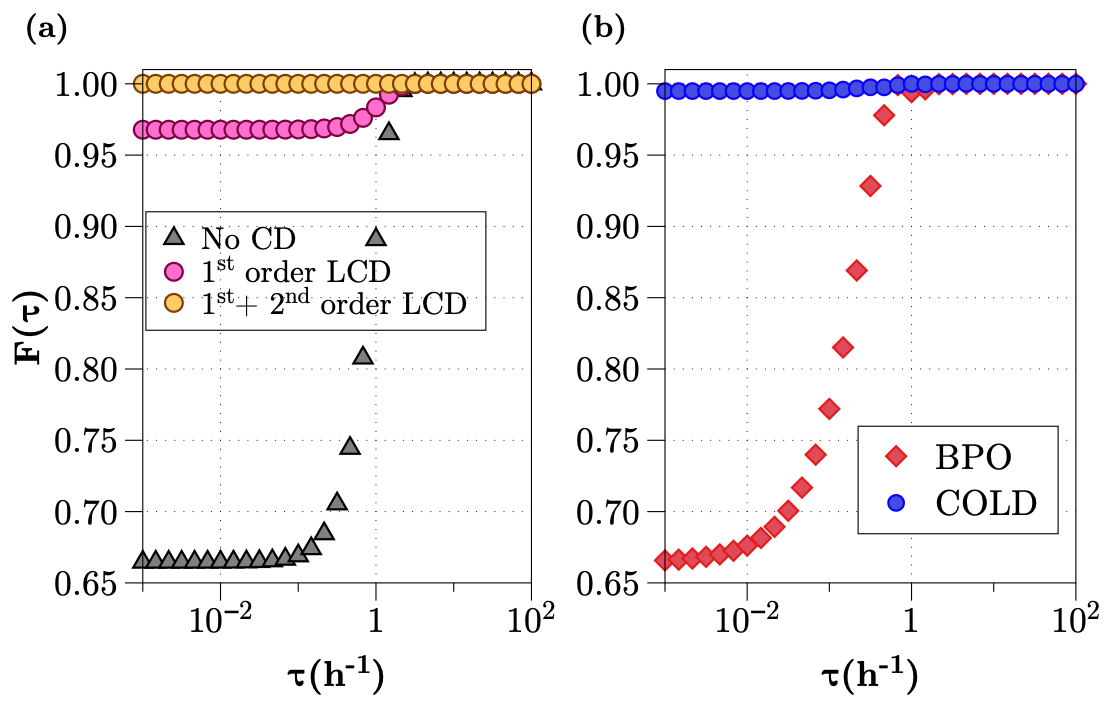}
     \caption{Optimisation of the annealing protocol for two spin Hamiltonian given by Eq.~\eqref{eq:Hanneal} for $h/J=2$.  (a) Final fidelities of the annealing protocol with triangles (black) representing the case where no CD is applied and circles showing the case of first-order LCD (pink) as well as the combination of first- and second-order LCD (orange). (b) Final fidelities achieved when using the optimal control method BPO (red diamonds) and the new approach of COLD (blue circles), both with $N_k=1$.}\label{fig:TwoSpin}
\end{figure}

We now consider three distinct cases in this two spin quantum annealing example: no LCD, first-order LCD, and second-order LCD. The fidelity of the final state for each case over a wide range of driving times is shown in Fig.~\ref{fig:TwoSpin}(a), with an easily distinguishable advantage in the case of LCD. The final fidelity where no LCD is implemented decreases rapidly as the ramp times are made short, with the system getting stuck in its initial state.  On the contrary,  first-order LCD retains good final state fidelities into short times, as the driving Hamiltonian becomes that of only the LCD term. The second-order LCD then gives unit fidelity, in agreement with previous observations \cite{claeys_floquet-engineering_2019}, as for a two spin Hamiltonian the highest order corrections are that including two spin terms.

We now add an optimisable term, as described in Sec.~\ref{sec:OptCont}, so that the new Hamiltonian reads:
\begin{equation}\label{eq:HannealOpt}
H_\beta(t) = H_0(t) + \sum_{k=1}^{N_k} \beta^k \sin (\pi k t / \tau) \sum_i \sigma_i^z,
\end{equation}
\noindent with $N_k$ the number of optimisation coefficients $\betabb$, and $\beta^k \in \betabb$ the coefficient of the $k$th frequency of the control function. Note that we consider
\begin{equation}
f(t,\betabb) =  \sum_{k=1}^{N_k} \beta^k \sin (\pi k t / \tau) =  \sum_{k=1}^{N_k} \beta^k \sin (\pi k g(\lambda)),
\end{equation}
with
\begin{equation}\label{eq:lambda}
g(\lambda) = \frac{2}{\pi}\arcsin\left(\sqrt{\frac{2}{\pi} \arcsin\left(\sqrt{\lambda}\right)}\right).
\end{equation}
The form of the additional control field fulfils the requirement that the boundary conditions are $H(0) = H_0(0)$ and $H(\tau) = H_0(\tau)$. Note that Numerically optimising the $\beta^k$ for the best final state fidelity \emph{without} adding LCD terms results in the BPO method introduced in Sec.~\ref{sec:OptCont}. We show the results of BPO in Fig.~\ref{fig:TwoSpin}(b), where it is observed that BPO gives better results than the case of no LCD in Fig.~\ref{fig:TwoSpin}(a). However, for short times the BPO approach still results in the system getting stuck in the initial state.

\begin{figure*}[t]
	\includegraphics[width=0.9\linewidth]{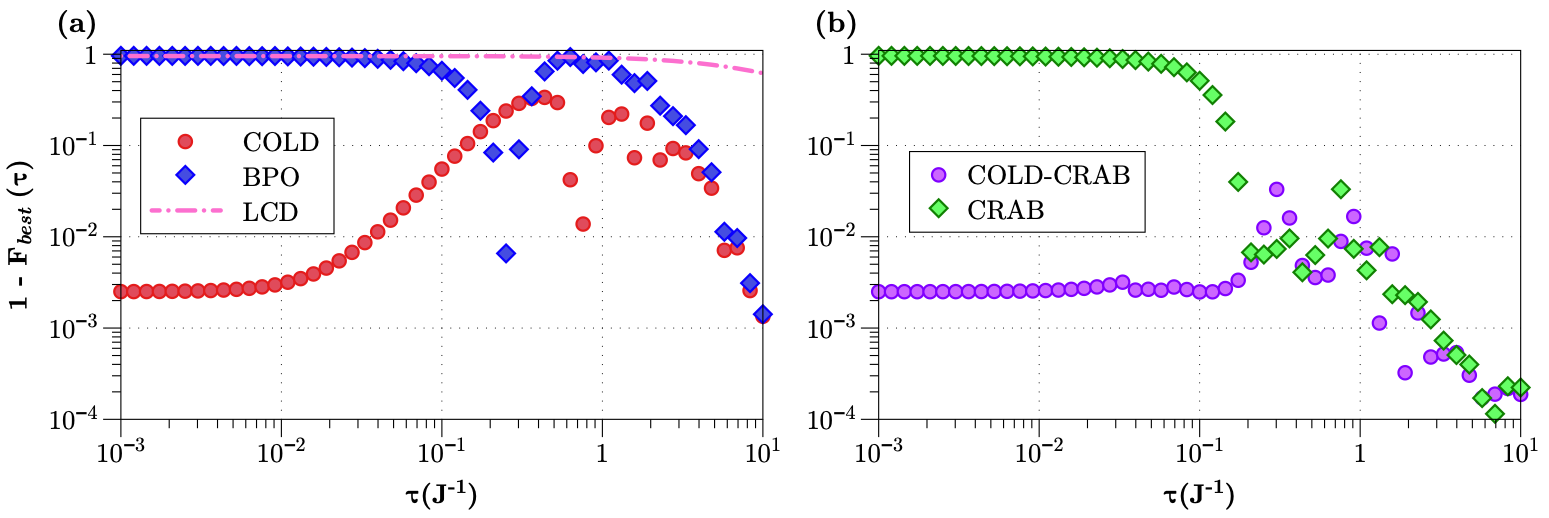}
     \caption{Optimisation of the annealing protocol for the Ising model given by Eq.~\eqref{eq:h0_ising} for $N=5$ spins. (a) A comparison of final state fidelities for different driving times using the optimal control technique BPO (blue diamonds), first-order LCD (pink dash-dot line) and COLD (red circles).  The same is shown in (b) for CRAB (green diamonds) and COLD-CRAB (purple circles). CD enhanced techniques (COLD and COLD-CRAB) introduced in this work show a clear convergence to good fidelities at short driving times.  All results are for the best (lowest) fidelity achieved over $500$ optimisations.}\label{fig:IsingUnconstrained}
\end{figure*}

Finally we present and compare the results of the new method, COLD. In this case the Hamiltonian before adding LCD terms is given by Eq.~\eqref{eq:HannealOpt} and the coefficient of the first-order LCD is
\begin{equation}
\alpha = -\frac{h(h+f(\lambda, \betabb)) + h\frac{\lambda}{\dot{\lambda}} \dot{f}(\lambda, \betabb)}{4(h\lambda)^2 + (h+f(\lambda, \betabb))^2 + 4J^2}.
\end{equation}
Note that the optimisation of the additional control field also feeds into the coefficient of the adiabatic gauge potential during the dynamics as discussed in Sec.~\ref{sec:OLCD}.  The results of the COLD approach for this two spin annealing protocol are shown in Fig.~\ref{fig:TwoSpin}(b), where we observe an improvement of the final state fidelity beyond what is possible with first-order LCD alone in Fig.~\ref{fig:TwoSpin}(a).  In this example, LCD alone reaches a final state fidelity of $1-F=3\%$ at short times, however COLD improves this error in the final state to $1-F=0.005\%$. This is due to the extended family of dynamical Hamiltonians in Eq.~\eqref{eq:expandedH} owing to the addition of an optimisable control field.  This result shows that COLD can provide an advantageous alternative to the addition of higher-order LCD which may be experimentally impractical.

We have found that COLD performs better than LCD of the same order or BPO when the system dynamics are calculated numerically.  This does not, however, imply anything about the performance of COLD in more complex scenarios, like in the case of an unknown target ground state.  In that case the fidelity is a poor optimisation metric. There is, however,  a way to come to the same conclusions as those presented in Fig.~\ref{fig:TwoSpin} without the need to compute the dynamics exactly.  We can do this by first using a guess for the COLD protocol to find the approximate adiabatic gauge potential and then minimising the integral of the norm of the second-order correction to the adiabatic gauge potential along the path.  Note, that the ground state can be in turn obtained through first order COLD, so there is no need to diagonalize the Hamiltonian. This integral should be small if COLD has implemented a dynamical Hamiltonian that makes the first-order adiabatic gauge potential the leading term.  It is effectively a measure of the error of COLD and can be given by
\begin{equation}
\begin{aligned}
\mathcal{I}_1(\Gamma) = \int_0^\tau dt^\prime \Big[& \bra{\psi_g(t^\prime)} \Gamma^2(t^\prime) \ket{\psi_g(t^\prime)}  \\ &- (\bra{\psi_g(t^\prime)} \Gamma(t^\prime) \ket{\psi_g(t^\prime)})^2\Big]^{1/2},
\end{aligned}
\end{equation}
\noindent with $\ket{\psi_g(t)}$ the instantaneous ground state along the path and 
\begin{equation}
\Gamma(t) = \gamma(t) \left( \sigma_1^y \sigma_2^x + \sigma_1^x \sigma_2^y \right),
\end{equation}
\noindent one of the second-order correction terms. In order to confirm this is the case, we compare the different paths -- COLD and LCD only -- in the two-spin example in order to determine if $\mathcal{I}_1$ is small for COLD. If $\mathcal{I}_1$ is small when compared to the same measure for lower-order LCD as $t\rightarrow 0$, then we know that COLD is enforcing a better dynamical Hamiltonian.  In the case of the two spin annealing protocol we find that as $t\rightarrow0$, $\mathcal{I}_1\rightarrow 0.04$ for COLD and $\mathcal{I}_1\rightarrow 0.2$ for LCD, showing that COLD is minimising the second-order correction along the path.  A simpler integral 
\begin{equation}\label{eq:integral2}
\mathcal{I}_2(\gamma) = \int_0^\tau dt^\prime |\gamma(t^\prime)|,
\end{equation}
also reflects this correction in this two spin example, with $\mathcal{I}_2 \rightarrow 0.03$ for COLD and $\mathcal{I}_2 \rightarrow 0.1$ for LCD as $t\rightarrow0$. This is particularly useful in more complex scenarios as $\mathcal{I}_2$ is relatively straight-forward to calculate, as we will demonstrate in the next section.  We also observe the reduction of the corresponding integrals of the $(\sigma^y_1\sigma^z_2 + \sigma^z_1\sigma^y_2)$ term of the second-order LCD. By minimising these integrals, it is possible to extend the COLD approach to more complex scenarios, including where the exact calculation of the dynamics is not possible.

\section{1D Ising Model} \label{sec:1dIsing}

In this section we apply COLD for state preparation on a 1D Ising spin chain in the presence of a transverse and longitudinal field. We consider an annealing protocol where the aim is to prepare the ground state across the Ising phase transition. The annealing Hamiltonian is given by
\begin{align}\label{eq:h0_ising}
	\begin{split}
    H_{0}(t) &= - J \sum_{j}^{N-1} \sigma^z_j\sigma_{j+1}^z + Z_0\sum_j^N \sigma_j^z \\ &+ \lambda(t) X_f \sum_j^N \sigma_j^x,
    \end{split}
\end{align}
where $Z_0$ is a small offset parameter to break ground state degeneracies and $X_f$ is the final x-field strength. Note, the breaking of the ground state degeneracies is not a requirement but allows for easier consideration of the adiabatic path. As before, $\lambda(t)$ is a scaling function that has the boundary conditions $\lambda(0) = 0$ and $\lambda(\tau) = 1$, with $\tau$ the driving time. This means we start from the ground state of all spins up and drive across the quantum phase transition to the ground state which is a superposition of all basis states. We again take the scaling function to be given by Eq.~\eqref{eq:Scalingfunc}.  In this example, we use $X_f = 10J$ and $Z_0 = 0.02J$.

For the Hamiltonian of  Eq.~\eqref{eq:h0_ising}, the LCD to first and second order is well known, as the wave functions are entirely real. We take the first-order adiabatic gauge potential to be given by
\begin{equation}\label{eq:LCD1}
\mathcal{A}(\lambda) = \alpha \sum_{j}^N\sigma_j^y,
\end{equation}
\noindent where the coefficients for the general periodic spin chain of Eq.~\eqref{eq:h0_ising} are
\begin{align} \label{eq:alphas}
    \alpha(\lambda) = \frac{1}{2} \frac{Z_0 X_f}{Z_0^2 + \lambda^2 X_f^2 + 2J^2}.
\end{align}
Note that the quoted $\alpha$ above is technically for a periodic or infinite size system, with $J^2 \rightarrow J^2(1-1/N)$ for a finite system. However, we find that the inclusion of the factor for the finite system sizes we consider only changes the final achieved converged fidelities at short times by $\sim 10^{-6}\%$. The second-order adiabatic gauge potential is of the form
\begin{equation}
\begin{aligned}
 \mathcal{A}^{(2)}(\lambda) =& \alpha \sum_{j} \sigma_j^y + \gamma  \sum_{j} (\sigma_j^x \sigma_{j+1}^y + \sigma_j^y \sigma_{j+1}^x) \\ & +  \zeta \sum_{j} (\sigma_j^z \sigma_{j+1}^y + \sigma_j^y \sigma_{j+1}^z) ,
 \end{aligned} \label{eq:SecondLCD}
\end{equation}
with the coefficients $\alpha$, $\gamma$ and $\zeta$ again obtained by minimising the action given by Eq.~\eqref{eq:actionCD} and solving the coupled set of equations numerically (see Appendix \ref{app:derivation} for a detailed derivation).

\begin{figure}[t]
	\includegraphics[width=0.98\linewidth]{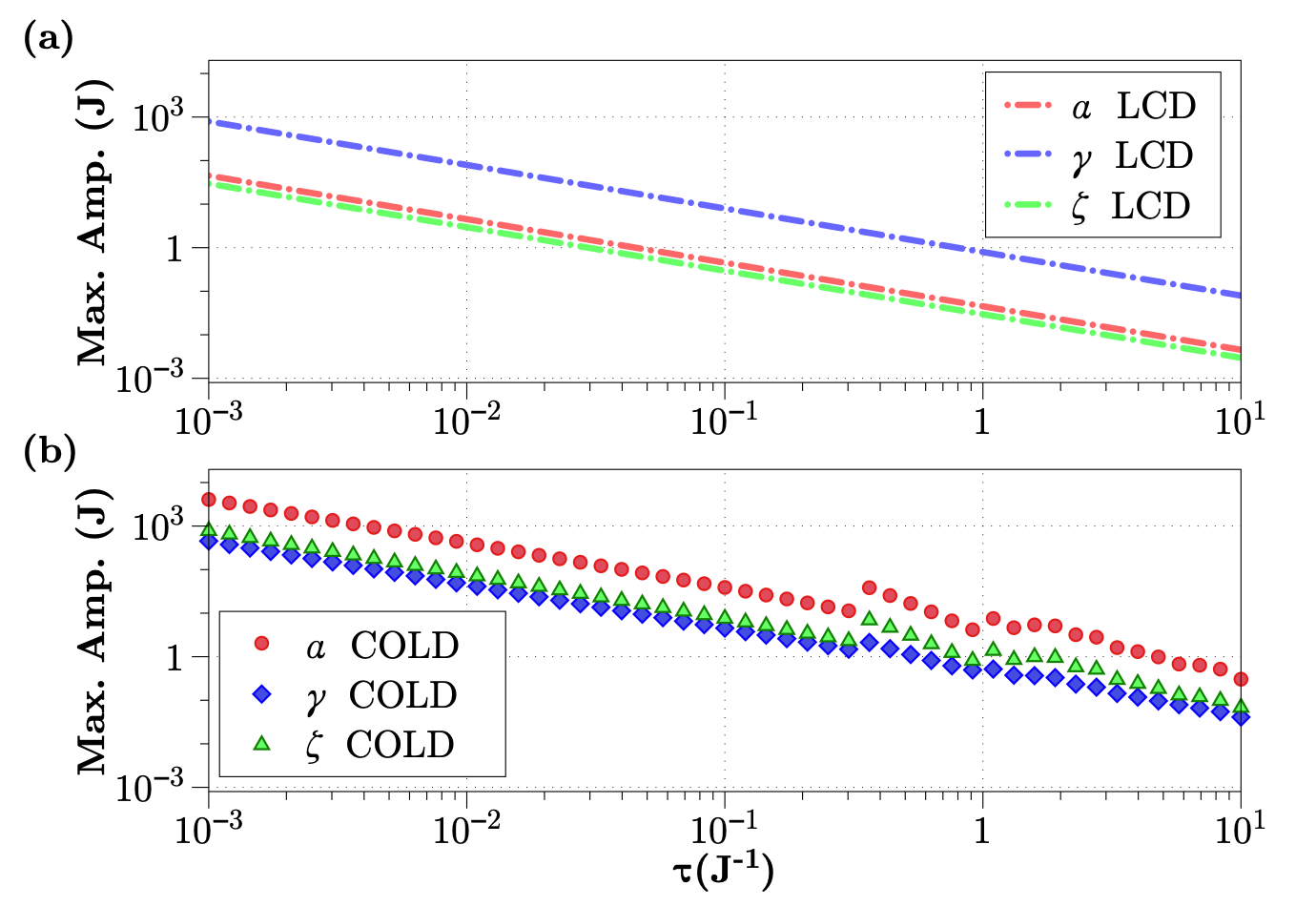}
     \caption{Maximum amplitudes of CD terms in the Ising model annealing protocol for (a) first- and second-order LCD only with no additional optimal control fields and (b) the COLD approach optimised for the best final state fidelity implementing first-order LCD as shown in Fig.~\ref{fig:IsingUnconstrained} (a). The plot shows the maximum amplitude at each driving time for the first-order $\alpha$ (red circles) and the two second-order terms $\gamma$ (blue diamonds) and $\zeta$ (green triangles) as given in Eq.~\eqref{eq:SecondLCD} (although the second-order LCD is not actually implemented in COLD). An inversion in the strength of the second-order and first-order LCD terms for (a) no additional optimal control fields and (b) the addition of optimal control fields shows that COLD implements a dynamical Hamiltonain which is favourable for the applied order of LCD (first-order in this case).}\label{fig:MaxAmp}
\end{figure}

In this example, optimal control is implemented by introducing an additional driving field so that the dynamical Hamiltonian is given by 
\begin{align}\label{eq:h_beta}
    H_{\beta}(t, \betabb) = H_0(t) + \sum_j f(t, \betabb)\sigma^z_j
\end{align}
with $\betabb$ being the terms to optimise over. We take our additional terms to again respect the boundary conditions $f(t=0, \betabb) = 0$ and $f(t=\tau,\betabb) = 0$, meaning a natural choice is
\begin{align}\label{eq:optimsable_1}
   f(t,\betabb) = \sum_k^{N_k} \beta^k \sin(\omega_k t / \tau) =  \sum_k^{N_k} \beta^k \sin(\omega_k g(\lambda)),
\end{align}
with $\omega_k = 2\pi k$ the $k$th principal frequency and $g(\lambda)$ given by Eq.~\eqref{eq:lambda}.  To implement the CRAB algorithm discussed in Sec.~\ref{sec:OptCont}, we will use $k \rightarrow k(1+r_k)$ instead with $r_k$ drawn from a uniform random distribution $r_k \in [-0.5,0.5]$.  Note that there is a strong distinction between the CRAB, which is an established optimal control method in its own right and our own version COLD-CRAB, which includes an LCD term along with the optimal control field in the Hamiltonian.  To be more precise, the COLD-CRAB Hamiltonian can be expressed as:
\begin{equation}\label{eq:cold_crab}
\begin{split}
 H_{\rm CC}(t,\betabb , \boldsymbol r) &= H_0(t) +  \alpha(t, \betabb, \boldsymbol r) \sum_{j}^N\sigma_j^y \\ &+ \sum_j f(t, \betabb, \boldsymbol r)\sigma^z_j,
 \end{split}
\end{equation}
where for each optimisable parameter $\beta^k$ associated with a $k^th$ principal frequency we also assign a random value $r_k \in \boldsymbol r$ as described earlier.  Note that the dependence on $\boldsymbol r$ is inherited by the LCD drive term $\alpha$, since it is a function of $f(t,\betabb, \boldsymbol r)$.

\begin{figure}[t]
	\includegraphics[width=0.98\linewidth]{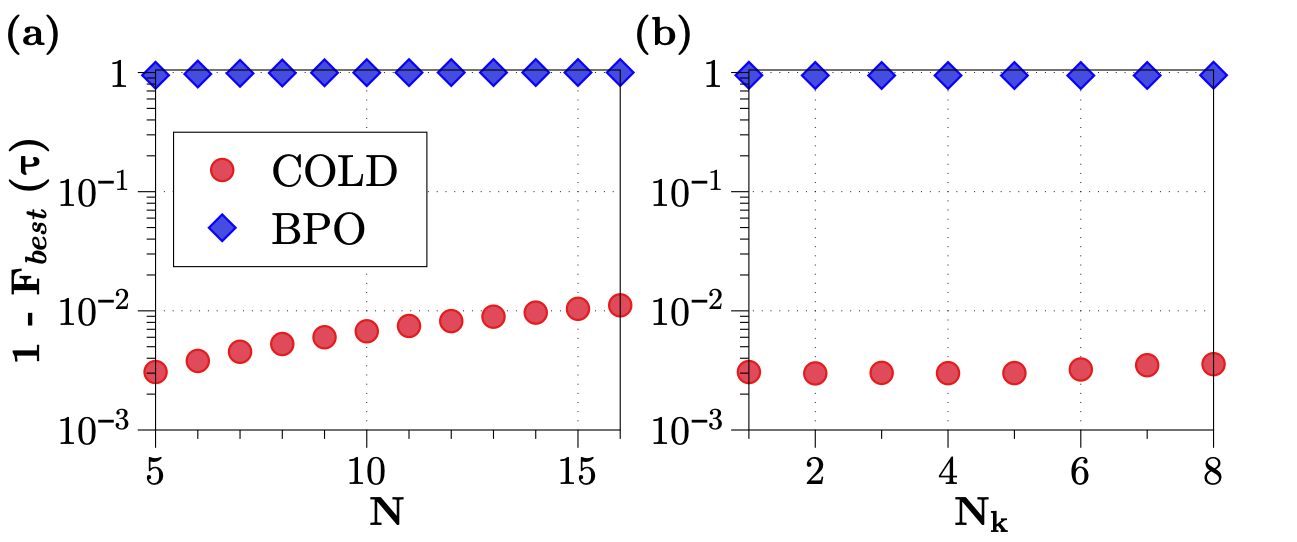}
     \caption{Scaling of fidelities in the annealing protocol for the Ising model with (a) system size $N$ and (b) optimisation parameters $N_k$ at driving time $\tau=10^{-2}J^{-1}$. Plots show a comparison between BPO (blue diamonds) and COLD (red circles).  In (a) we see that the COLD fidelity decreases as a function of $N$ but remains quite high when compared to BPO while (b) shows the non-existent improvement for both BPO and COLD with an increasing number of parameters in the $N=5$ spin case. Once again, plotted best fidelities are obtained across 500 optimisations.}\label{fig:ScalingN}
\end{figure}

As before, we choose the first order adiabatic gauge potential given by Eq.~\eqref{eq:LCD1} and find that the coefficients are
\begin{align}
    \alpha(\lambda,\betabb) = \frac{X_f}{2} \frac{(Z_0 + f(\lambda,\betabb)) - \lambda\dot{f}(\lambda,\betabb)/\dot{\lambda}}{(Z_0 +f(\lambda,\betabb))^2 + \lambda^2 X_f^2 + 2 J^2}.
\end{align}
Note, with the introduction of the additional control fields $f$ it is possible for $\alpha$ to be non-zero at the start or end of the protocol, as $\dot{f}$ is not fixed to be zero. However, this can be enforced by a suitable choice of the additional control field, we will consider replacing $\alpha \rightarrow S(\lambda)\alpha$ where $S(\lambda)$ is a scaling function that tends to zero as $\lambda \rightarrow 0$ and $\lambda \rightarrow 1$. We find that the scaling function only has a minimal effect on the final fidelities observed. This issue could also be resolved by a suitable choice of $f$, with our example drive being an extreme case as $\dot{f}$ is maximal at the boundaries of the protocol.  Note that this issue is present in LCD as much as in COLD and we have chosen to highlight it here as it may become a concern in an experimental setting where a discontinuous drive is simply impossible at the beginning and end of a protocol. The suitable choice of the form of $f$ in a given example is a problem we will leave for future work, with our focus being on the introduction of the COLD protocol.

We first compare the final state fidelity when using COLD versus BPO as shown in Fig.~\ref{fig:IsingUnconstrained}(a) for different driving times in a system of $N=5$ spins and a single $N_k=1$ optimisation coefficient. As expected, at long timescales the two methods agree as we approach the adiabatic limit of the dynamics. However, at shorter time scales the difference in behaviour is dramatic. We observe that the BPO approach fails in the case of very fast driving as the state gets stuck in the initial state but the COLD approach converges to $1-F \sim 10^{-3}$.  We note that the advantage achieved by COLD is not due to the introduction of first-order LCD terms alone, as in Fig.~\ref{fig:IsingUnconstrained}(a) we see that this will result in $F=0.0440$ for $\tau=10^{-3}J^{-1}$. COLD is instead achieving this by making the LCD term dominant for the dynamical Hamiltonian through the additional control fields.  

To confirm this, we plot the maximum amplitudes of both the first- and second-order adiabatic gauge potentials in Fig.~\ref{fig:MaxAmp}, where Fig.~\ref{fig:MaxAmp}(a) shows the case of no optimisation and Fig.~\ref{fig:MaxAmp}(b) the case of applying COLD. We can see that without COLD the second-order $(\sigma^x_j\sigma^y_{j+1} + \sigma^y_j\sigma^x_{j+1})$ corrections to the LCD are far larger than the first-order,  resulting in the small final state fidelities when only first-order LCD is implemented. In the case of COLD, this relationship reverses and the first-order LCD terms dominate the dynamics.  This gives us some indication that one way to optimise the control pulse may be a minimisation of higher order LCD terms, which we explore further in Sec.~\ref{sec:minimisation}.

\begin{figure*}[t]
	\includegraphics[width=0.9\textwidth]{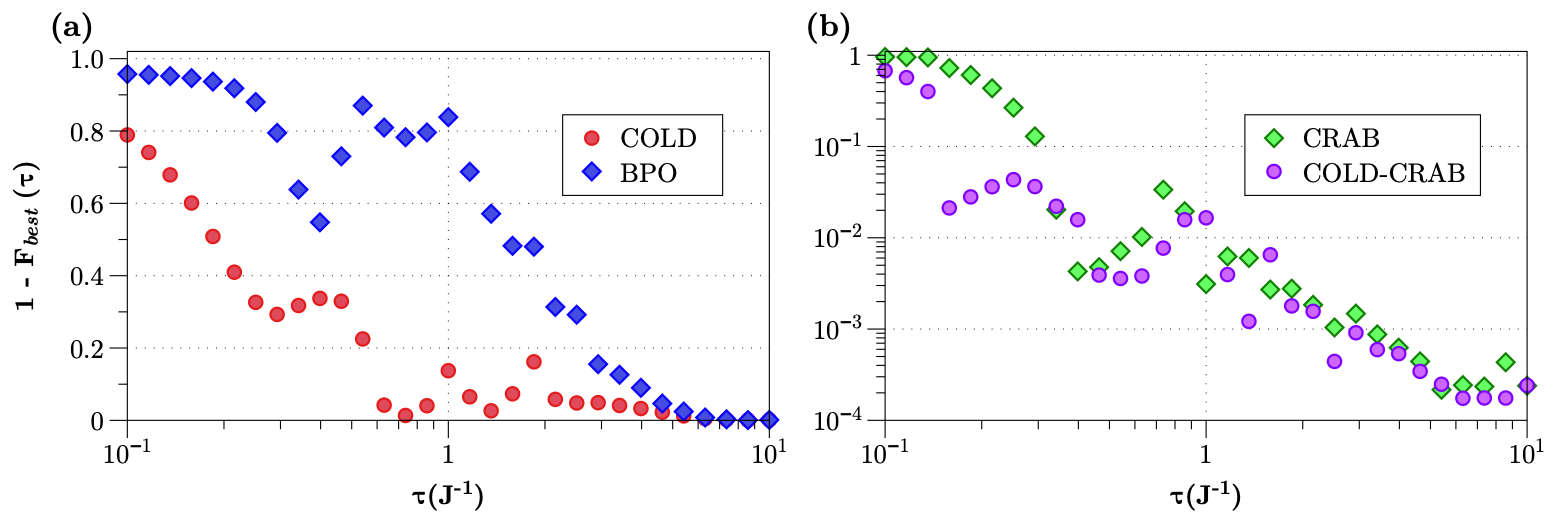}
    \caption{Optimisation of the \emph{constrained} annealing protocol for the Ising model for $N=5$ spins with a maximum amplitude limit on each term in the Hamiltonian of Eq.~\eqref{eq:h0_ising} of $10J$.  (a) Shows a comparison between BPO (blue diamonds) and COLD (red circles) which both give much lower fidelities than in the unconstrained case in Fig.~\ref{fig:IsingUnconstrained}, although COLD persists in giving better results. In (b) the comparison is between CRAB (green diamonds) and COLD-CRAB (purple circles) which show orders of magnitude better fidelities than those in (a), with COLD-CRAB eking out higher fidelities at short driving times. The plotted best results are obtained from 200 optimisations for each method.}\label{fig:IsingcConstrained}
\end{figure*}

We find that the results of BPO and COLD at short driving times are stable against increasing system size, as shown for $\tau = 10^{-2}J^{-1}$ in Fig.~\ref{fig:ScalingN}(a), with only a small decrease in final state fidelity for larger systems with COLD. Similarly,  increasing the number of optimisation coefficients $N_k$ results in little improvement in the values obtained at short times for this example, as shown in Fig.~\ref{fig:ScalingN}(b).  It is possible that in more complex systems, more optimisation coefficients will be needed to gain a larger advantage. We also note that by increasing the number of coefficients, we are increasing the complexity of the cost function landscape to be explored by the minimisation procedure, hence leading to slightly worse final fidelities This can mean that alternative approaches than the Powell minimisation used so far, \@e.g.~that of CRAB, could be better suited to probing the cost function for high $N_k$. We also note that this lack of improvement in the results is likely the consequence of the form of the control field given by Eq.~\eqref{eq:optimsable_1} rather than due to a failure of the optimiser in the face of a complex parameter space. We find that the parameter space is relatively smooth in the case of $N_k = 1,2,3$ and a better solution for this form of control field does not exist. 

We now consider the combined method of COLD-CRAB for this annealing example as shown in Fig.~\ref{fig:IsingUnconstrained}(b). We point out that with our application of CRAB in this scenario we are not enforcing $\betabb$ to be zero at the start and end of the dynamics, allowing for their to be a tuning of the $z$-field offset. This is consistent between CRAB and COLD-CRAB and therefore does not influence our comparison of the two. First, it is important to note that CRAB alone results in a overall speedup of the dynamics for a high final state fidelity $1-F\sim 10^{-3}$ at long time-scales. However, CRAB still suffers from getting stuck in the initial state at fast driving times and the final state fidelity again tends to zero. This is not the case for COLD-CRAB, which converges to large final state fidelities $1-F \sim 10^{-3}$ at short driving times $\tau \leq 10^{-1}J^{-1}$. Note that the difference between the convergence to final state fidelities are only marginally different between COLD and COLD-CRAB at longer times,  but at short times COLD-CRAB performs a lot better. Further improvement could be gained by combining COLD with more advanced versions of CRAB or other optimal control methods.

As shown in Fig.~\ref{fig:MaxAmp}, the amplitude of the driving required to achieve the fidelities discussed so far scales with the driving time. Practical scenarios will necessarily place limits both on achievable driving times and the maximum amplitude of any term that is being driven.  However, the scaling of the drivings shown do not mean that everything diverges in the limit of $\tau \rightarrow 0$. To see this we can first write the Scr\"odinger equation for COLD as
\begin{equation}
i \hbar d_t \ket{\psi}=\left(H_\beta+\dot\lambda \mathcal{A}_\lambda\right) \ket{\psi},
\end{equation}
we then divide through by $\dot{\lambda}$ to get
\begin{equation}
i \hbar d_\lambda \ket{\psi}=\left(\frac{H_\beta}{\dot \lambda}+\mathcal{A}_\lambda\right) \ket{\psi},
\end{equation}
in the limit of $\tau \rightarrow 0$ then $\dot\lambda\to \infty$ to result in the Hamiltonian term disappearing,  or in other words, we turn off the Hamiltonian. We then only drive the system in the $\tau \to 0$ limit with the COLD or LCD driving term:
\begin{equation} \label{eq:OnlyA}
i \hbar d_\lambda \ket{\psi}=\mathcal{A}_\lambda \ket{\psi}.
\end{equation}
In this limit then $\lambda$ plays the role of time, and this could then be implemented in a practical scenario in finite time as it corresponds to some manipulation of the couplings in the system. This renormalised time cannot then be infinitesimally short if the couplings are bounded but we have shown that the protocol does not diverge as $\tau \rightarrow 0$. In the case of a spin chain, evolving under Eq.~\eqref{eq:OnlyA} is effectively to first order in LCD implementing independent single spin rotations along the chain, and COLD can be easily applied \cite{caneva2011speeding, murphy2010communication}.

If it is not possible to switch off the Hamiltonian as discussed above then as an alternative we can implement COLD with experimental constraints accounted for directly in the optimal control minimisation. We consider an extreme example of constraints to show that even in this scenario COLD can provide an advantage and corresponding speed-up.  In the constrained case the annealing protocol remains that of Hamiltonian~\eqref{eq:h0_ising} but we choose to introduce a bound of $X_f$ on the maximum amplitude of all drivings. This makes it so that no optimal control or LCD term can go beyond the original amplitude of the $x$-field drive. We show the final state fidelities achieved for the constrained example in Fig.~\ref{fig:IsingcConstrained}. As can be seen in Fig.~\ref{fig:IsingcConstrained}(a), COLD provides a substantial improvement beyond what is achievable with BPO. BPO manages $F < 0.5$ for $\tau < 1J^{-1}$,  but COLD can reach final state fidelities $F\sim 0.9$ for $\tau < 1J^{-1}$. The real improvement, however, comes with the application of CRAB and COLD-CRAB. CRAB already improves the fidelities substantially, and would allow for a speed up in the annealing protocol but with COLD-CRAB the final state fidelities are even better, with $F\sim 0.99$ achievable when approaching $\tau\sim 0.1J^{-1}$. Signs are seen of the onset of the convergence to small values for COLD-CRAB in Fig.~\ref{fig:IsingcConstrained}(b) before the maximum amplitude required becomes too large and the short time results tend towards zero fidelity and the state being stuck again. With this example and the discussion on implementation via turning off the Hamiltonian, we have shown that COLD is capable of delivering improvements beyond other schemes even for practical problems with strict and rather extreme constraints.

\section{Transport in a Synthetic Lattice}\label{sec:lattice}

\begin{figure*}[t]
	\includegraphics[width=0.98\textwidth]{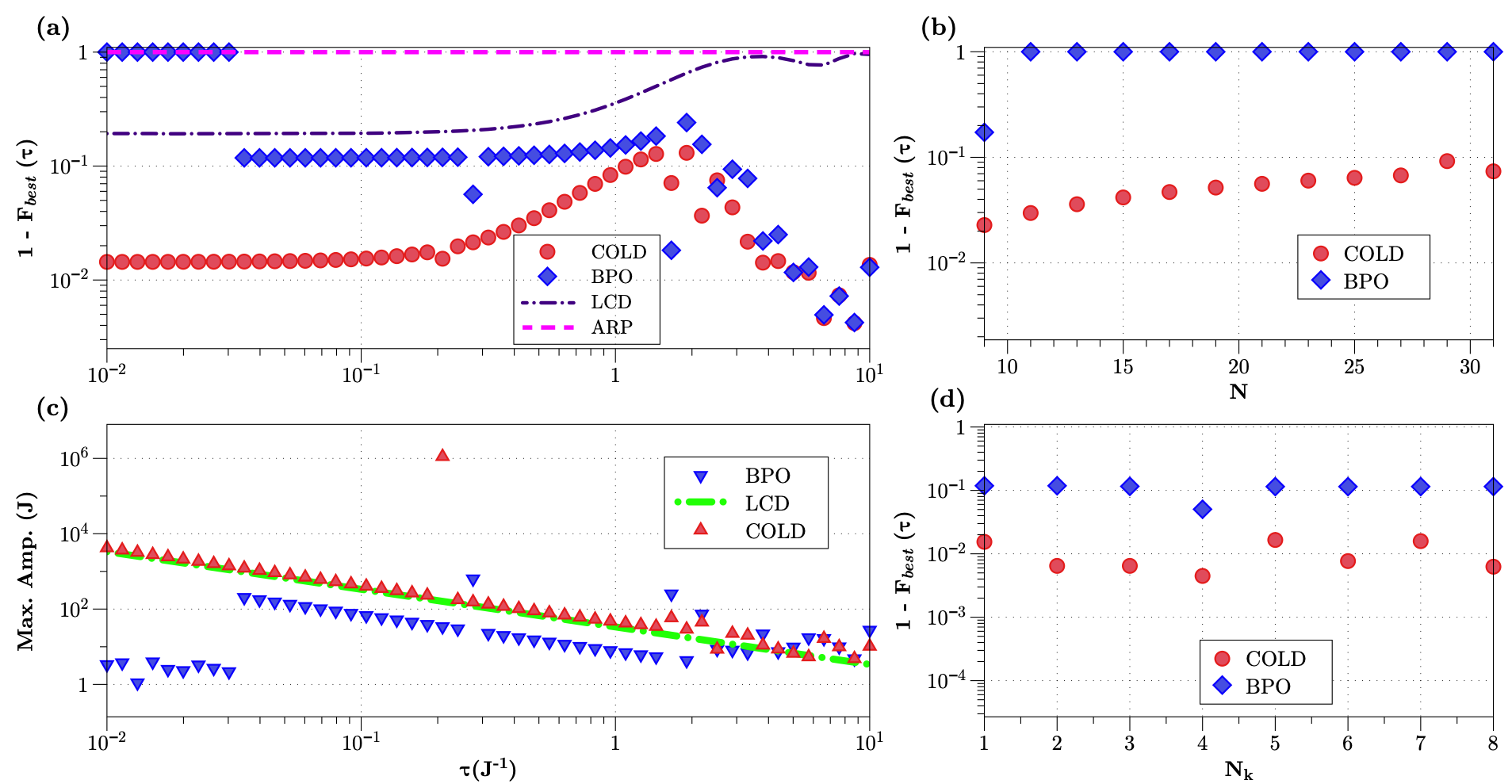}
    \caption{Optimisation of state transfer in a synthetic lattice.  In (a) we compare the fidelities obtained via the bare ARP protocol (pink dashed line) and first-order LCD previously implemented in Ref.~\cite{meier_counterdiabatic_2020} (purple dash-dot line) to BPO (blue diamonds) and the COLD method (red circles).  (c) Maximum amplitude of the tunneling term at each driving time for LCD (green diamonds) as given by Eq.~\eqref{eq:tunneling} as well as COLD (red triangles) which includes additional control parameters as shown in Eq.~\eqref{eq:tunneling_opt} and BPO (blue triangles) which omits the modifications due to CD but retains the control terms $\betabb$.  In both (a) and (c) we simulate $N = 7$ lattice sites and use $N_k = 1$ parameter for optimisation of BPO and COLD.  (b) Scaling of fidelities with increasing number of lattice sites (where $N_k = 1$) for both COLD (red circles) and BPO (blue diamonds) noting that the latter performs very poorly for $N > 9$. (d) does the same for the number of parameters while keeping $N=7$, with the trend indicating that increasing $N_k$ does not lead to better fidelities in either the BPO or COLD case.  Note that both (b) and (d) are simulated for driving time $\tau = 0.5 J^{-1}$ and the best fidelities are obtained across 500 optimisations.}\label{fig:Synthetic}
\end{figure*}

The efficient transfer of states between opposite ends of a lattice could have future applications in the settings of quantum computation and simulation due to its promise of efficient transport of information \cite{lang2017}. This objective is often tackled in the setting of ultracold atoms in optical lattices. While the problem can be tuned to be a single-particle system and the analytical solutions of the corresponding instantaneous Schr\"odinger equation are known \cite{Hatsugai1993,Hugel2014} even for a finite system \cite{Duncan2018exact}, efficient evolution for state transfer is not straight-forward.  This is due to the fact that the majority of the states are delocalised across the lattice, meaning that the $\ket{\psi}\bra{\psi}$ terms of the CD Hamiltonian of Eq.~\eqref{eq:counterdiabatic} are global in reach.  It is normal to consider this system in the tight-binding limit where the implementation of global terms is not straightforward. Such terms can be generated via the interactions of the atoms with cavity modes \cite{landig2016,keller2017} or from dipolar interactions \cite{baranov2002,menotti2008,trefzger2011}. However, it would be ambitious to expect this control to be general enough to implement the CD Hamiltonian of the exact solutions.  This is one of the reasons that LCD has been pursued in this setting. 

Recently, LCD has been successfully applied to improve an adiabatic rapid passage (ARP) protocol for population transfer across a synthetic lattice \cite{meier_counterdiabatic_2020}. In this realisation, population transfer was achieved in a synthetic tight-binding lattice of laser coupled atomic momentum states. We will consider the same problem as in Ref.~\cite{meier_counterdiabatic_2020} but with the improvement that can be gained by COLD. This system is described by the Hamiltonian
\begin{align}\label{eq:Hlattice}
\begin{split}
    H_0(t) &= - \sum_n J_n(t)(c_n^{\dag}c_{n+1} + H.c.) \\ &+ \sum_n V_n(t) c_n^{\dag}c_n,
\end{split}
\end{align}
where $J_n(t)$ is the time-dependent tunnelling that describes the nearest-neighbour coupling, $V_n(t)$ is the on-site energy offset with respect to neighbouring sites and $c_n^{\dag}$($c_{n}$) is the creation(annihilation) operator on a given synthetic lattice site. In the ARP protocol, the population gets moved from one end of the lattice to the other by linearly ramping the lattice from a positive tilt to a negative tilt via
\begin{align}\label{eq:t_and_v}
    J_n(t) = J_0(1.1 - \lambda) = J_0\Big(0.1 + \frac{t}{\tau}\Big), \\
    V_n(t) = n V_0 2 (\lambda - 1/2) = nV_0\Big(1 - \frac{2t}{\tau}\Big),
\end{align}
where $V_0 = 4J_0$ is the initial site energy slope and $J_0$ is the characteristic tunnelling scale of the lattice. The scaling function in this case is given by
\begin{equation}
\lambda(t) = 1-\frac{t}{\tau}.
\end{equation}
In order to implement LCD as shown in Ref.~\cite{meier_counterdiabatic_2020}, the first order LCD can be accounted for by taking
\begin{align}\label{eq:tunneling}
    J_n(t) \rightarrow J_{n, \mathrm{CD}}(t) e^{-i\phi_{n, \mathrm{CD}}(t)},
\end{align}
where
\begin{align}\label{eq:t_phi_cd}
    J_{n, \mathrm{CD}}(t) = \sqrt{J_n(t)^2 + (\alpha_n(t)/\tau)^2}, \\
    \phi_{n, \mathrm{CD}}(t)  = \arctan\left(-\frac{J_n(t)\tau}{\alpha_n(t)}\right),
\end{align}
and $\alpha_n(t)$ is the CD terms which can be found by solving a set of linear equations
\begin{align}
    \begin{split}
        &-3(J_n J_{n+1})\alpha_{n+1} + (J_{n-1}^2 + 4J^2_n + J_{n+1}^2)\alpha_j \\ &- 3(J_n J_{n-1})\alpha_{n-1} + (V_{n+1} - V_n)^2 \alpha_n \\ &= -\partial_{\lambda}J_n (V_{n+1} - V_{n}).
    \end{split}
\end{align}
In order to implement COLD we include additional terms to the tunnelling of the lattice 
\begin{align}\label{eq:tunneling_opt}
    J_n(t) \rightarrow J_n(t, \betabb) = J_n(t) + f(t, \betabb),
\end{align}
which can then be incorporated into the forms of both $J_{n, CD}(t)$ and $\phi_{n, CD}(t)$. We again want the additional control terms to go to zero around the problem boundaries and a natural choice is the same as in the Ising spin chain example in Eq.~\eqref{eq:optimsable_1}. The parameters $\betabb$ are optimised as before by minimizing with respect to the fidelity of the final state, where the population has been fully transferred to the opposite lattice site. 

We first consider a system size of $N=7$ sites which was successfully experimentally probed in Ref.~\cite{meier_counterdiabatic_2020}, where final state fidelities of $0.75$ were achieved for $\tau = 1$ms with a final tunnelling strength of $J/\hbar = 1/2\pi kHz$ (equivalent to $\tau \sim 1 J^{-1}$ in our units). We initially confirm the breakdown of ARP in this setting for fast times, and the success of the LCD protocol at short times, as shown in Fig.~\ref{fig:Synthetic} (a) and found in Ref.~\cite{meier_counterdiabatic_2020}. Implementing BPO on its own manages to enhance the achievable fidelities at intermediate times of $\tau > 0.03 J^{-1}$. However, eventually, as observed in all scenarios in this work, BPO becomes stuck in the initial state at fast times, and the fidelity goes to zero. Implementing the newly introduced COLD protocol achieves an order of magnitude improvement in the fidelity over LCD. This is also plotted in Fig.~\ref{fig:Synthetic}(a) alongside previous results of ARP and first-order LCD.

One concern could be that COLD is achieving this improvement by simply pumping power into the tunnelling term, but as we can see in Fig.~\ref{fig:Synthetic}(c) the maximum amplitude of the tunnelling term tracks that of LCD. A key issue for experiments is the maximum amplitude achievable by a driving term and with this result we can stipulate that COLD is likely to be feasible in the same regimes as LCD in this synthetic lattice system. There is single outlier at intermediate times as indicated by the single point peaking in maximum amplitude in Fig.~\ref{fig:Synthetic}(c), this is the exception to the rule, where the optimisation has found a marginally higher fidelity (see the offset point in Fig.~\ref{fig:Synthetic}(a)) by pumping in power.

A large concern for state transfer techniques is the robustness of a protocol with respect to an increasing system size. We show the best achievable fidelities with increasing system size for both BPO and COLD in Fig.~\ref{fig:Synthetic}(c). While both protocols show a decreasing fidelity with system size as is to be expected, once again COLD does not suffer from getting stuck in the initial state. This is shown by the BPO fidelities going to unity for large systems in Fig.~\ref{fig:Synthetic}(c), and is the same mechanism for this as for the short driving times in Fig.~\ref{fig:Synthetic}(a).

\begin{figure*}[t]
	\includegraphics[width=0.9\textwidth]{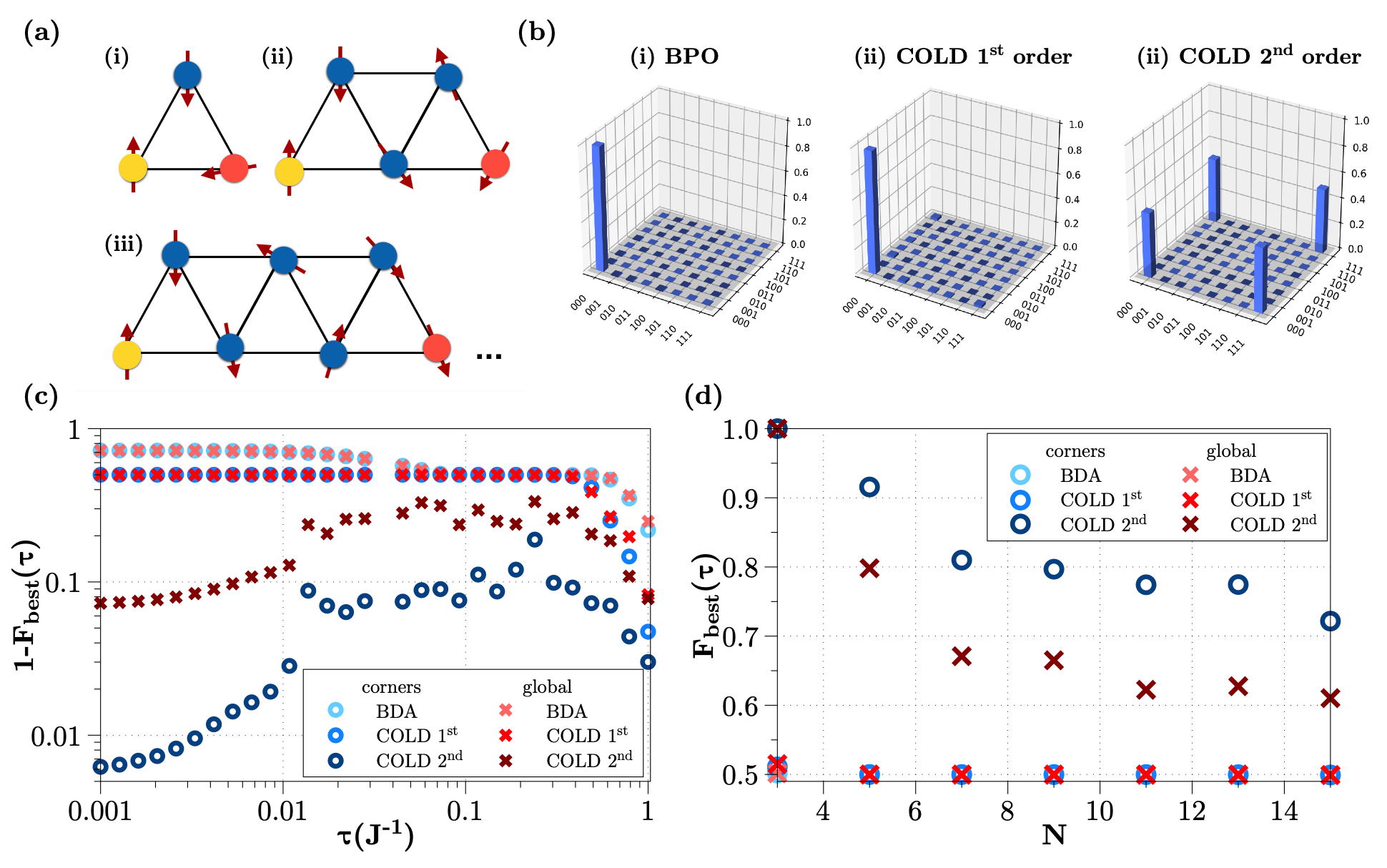}
    \caption{GHZ state preparation in systems of frustrated spins.  Spins are arranged in triangular formations as depicted in (a) for (i) 3, (ii) 5 and (iii)7 spins, with spins on the vertices and edges representing couplings.  In the case of corner optimisation,  three separate optimisable drives are applied: one for the yellow corner spin, one for the red corner spin and then a third drive for all of the blue spins in-between. (b) Density matrix plots of the final state of a 3 spin triangle after an evolution time $\tau = 0.1J^{-1}$ when optimised using (i) BDA, (ii) $1^{st}$ order COLD and (iii) $2^{nd}$ order COLD and corner optimisation. (c) Final fidelities of the GHZ state for the 5 spin configuration depicted in (a)(ii) for an optimised global drive (red crosses) and locally driven corner spins (blue rings). (d) Final fidelities  at driving time $\tau = 0.1J^{-1}$ for different systems sizes $N$. In the global case we use 10 total optimisable parameters with $N_k = 1$ drive and $N_m = 10$ time intervals. while in the corners case there are 30 total parameters as we increase to $N_k = 3$ separate drives. The plotted fidelities are the best results of 5 optimisations for each data point.}\label{fig:GHZ}
\end{figure*}

Another concern could be that BPO will beat COLD if enough parameters are allowed for the optimisation, i.e. if we increase $N_k$ enough. We observed no evidence of this for the Ising model example and we again do not observe this in this synthetic lattice example, as is shown in Fig.~\ref{fig:Synthetic}(d). Small improvements are made in the fidelities achieved with BPO and COLD for larger $N_k$ but this is not substantial. 

\section{GHZ state preparation}\label{sec:ghz}

As a final example we focus on the preparation of multi-partite Greenberger–Horne–Zeilinger (GHZ) \cite{greenberger1990bells} states:
\begin{align}
\ket{\rm GHZ} = \frac{1}{\sqrt{2}}(\ket{0}^{\otimes N} + \ket{1}^{\otimes N})
\end{align}
in a system of frustrated spins (see Fig.~\ref{fig:GHZ}(a)).   We start out with a system of all spins pointing down and drive a bare Hamiltonian of the form:
\begin{align}\label{eq:Hghz}
\begin{split}
    H_0(t) &= -J \Big( \sum_j^{N-1} \sigma^z_j \sigma^z_{j+1} + \sum_j^{N-2} \sigma^z_j \sigma^z_{j+2} \Big) \\ & - h (1 - \lambda(t)) \sum_j^N (\sigma_j^x + \sigma_j^z),
\end{split}
\end{align}
where $J=1$ and $h = 10J$ with the same $\lambda(t)$ as used previously, given by Eq.~\eqref{eq:Scalingfunc}.  The form of the LCD to first and second order is the same as in the case of the Ising spin chain (see Eq.~\eqref{eq:LCD1} and Eq.~\eqref{eq:SecondLCD}) with the couplings in the case of the second order now including the additional terms between spins $j$ and $j+2$.

\begin{figure*}[t]
	\includegraphics[width=0.98\textwidth]{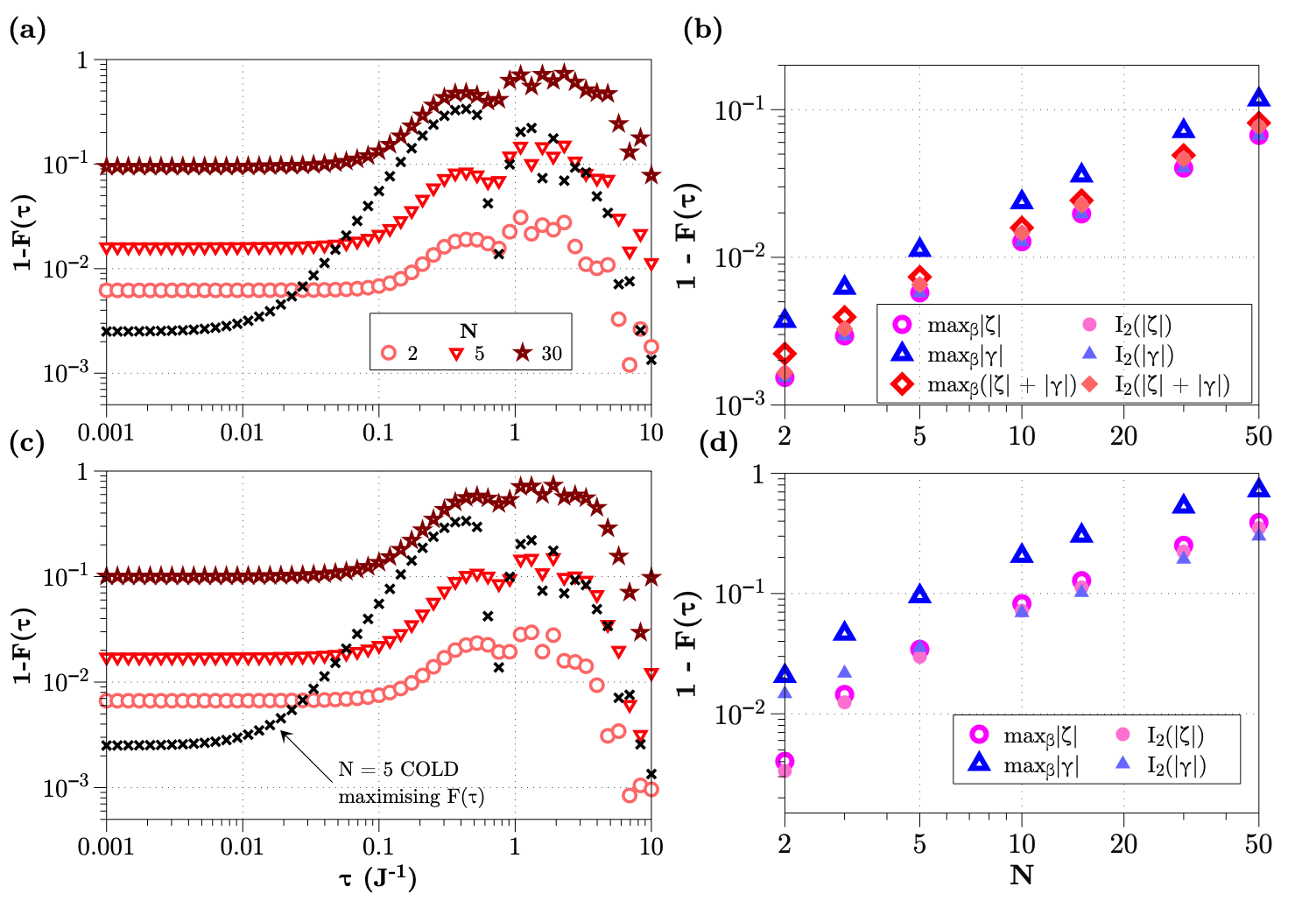}
    \caption{Optimisation of $\beta$ via minimising second-order LCD terms for the Ising model. In (a) we plot the final state fidelities after minimising the integral from Eq.~\eqref{eq:integral2} of the $|\zeta(t)|$ drive ($\mathcal{I}_2(|\zeta(t)|)$) and then apply the result at each driving time to different system sizes $N$. The results of minimising maximum amplitude instead $\rm max_\beta |\zeta(t)|$ are plotted in (c). The black crosses in both (a) and (c) plots are the results of optimising $\beta$ by maximising final state fidelity $F(\tau)$ and are the same as the red circle plot in Fig.~\ref{fig:IsingUnconstrained}(a).  Plotted in (b) and (d) are final state fidelities for $\tau = 0.1J^{-1}$ for different system sizes $N$ when optimising $\betabb$ either by minimising the maximum amplitude of a drive $\rm max_\beta(\cdot)$ or its integral $\mathcal{I}_2(\cdot)$. In (b) only first order COLD is applied post-optimisation while in (d) one of the second-order drives is also applied after minimising the other, \@e.g.~if $\rm max_\beta|\zeta|$ is minimised to determine the optimal $\betabb$, then both the first-order drive $\alpha$ and the other second-order drive $\gamma$ are applied.  For system sizes above N=10 we used ITensor\cite{itensor} MPS calculations which were converged with a truncation level of $10^{-14}$ per time step. at each site reaching a maximum bond dimension of $D = 4$. In all cases, a single optimisable parameter is used ($N_k = 1$).}\label{fig:Int_Min}
\end{figure*}

In order to explore the versatility of combining optimal control with LCD, for this example we design our optimal control drive according to the gradient ascent pulse engineering (GRAPE) method \cite{khaneja2005optimal}.  Our control coefficients $\beta^k$ are now discretised on a finite grid of $N_m$ time intervals $t_m$ with uniform step $\Delta t$  to obtain control sequences in which individual elements $\beta^{k,m}(t_m)$ are treated as continuous parameters
\begin{align} \label{eq:GRAPE_drive}
	\begin{split}
		f(t, \betabb) \rightarrow & \Big[ [f(\beta^{1,1},t_1), \cdots, f(\beta^{1,N_m},t_{N_m})],  \cdots, \\ &[ f(\beta^{N_k, 1},t_{1}), \cdots ,f(\beta^{N_k, N_m},t_{N_m})] \Big],
	\end{split}
\end{align}
where the total driving time $\tau = N_m \Delta t$ and $k$ is used to denote a localised drive for a subset of spins where $N_k$ is the total number of control pulses. As in the Ising model case, we take our optimal control Hamiltonian to be of the form in Eq.~\eqref{eq:h_beta} with each $k^{th}$ drive acting on the specified subset of spins with local $\sigma^z$ operators. At each time interval $t_m$ the $k^{th}$ control drive strength is calculated as:
\begin{align}
\begin{split}
		&f(\beta^{k,m}, t_m) \\
		&= \beta^{k,m} \tanh (\kappa \theta(t_m)) \tanh (- \kappa \theta(t_m - \tau)),
\end{split}
\end{align}
with $\theta(t) = \sin \frac{\pi t}{2 \tau}$ and $\kappa = 30$ an offset parameter used to control the shape of the drive. We use spline interpolation to calculate the derivatives of the control drive when they are required to obtain the LCD drives. The resulting function requires more parameters than the Fourier basis we chose to use in previous examples, however it also allows for more flexibility in the final shape of the drive. Furthermore, due the increased number of parameters and search space, instead of Powell optimisation as in previous examples we choose to instead implement dual annealing, which is a global optimiser and while computationally more costly, is far better in the case of a complex parameter space with multiple minima. 

Since such a preparation of GHZ states involves the generation of entanglement in a system that initially contains none, we expect that the first-order COLD may not contain the leading order of the counterdiabatic drive and thus not be as effective. For this reason, we include second-order COLD terms as given in Eq.~\eqref{eq:SecondLCD}.  We also explore the idea of using multiple control drives and localising them to parts of the system. Thus we implement both a global drive which is uniform across all spins as well as a `corner' evolution, in which three different optimisable drives are used: one each for the first and last spin in the lattice as well as one for all of the remaining spins. This is depicted in Fig.~\ref{fig:GHZ}(a), where different vertex (spin) colours represent different control pulses.

The results for a 5 spin system with control drives consisting of $N_m = 10$ time intervals are plotted in Fig.~\ref{fig:GHZ}(c), where we observe that first order COLD is indeed not particularly effective at short driving times and does not move the system out of its initial state (see density matrix plots in (b)), regardless of whether or not separate control is applied to the corner spins. This is very likely due to the fact that the local $\sigma^y$ terms are only a small contribution to the full counterdiabatic drive and thus we need to look to higher order LCD to see any improvements. This is exactly what the results indicate, as second order COLD shows a five-fold improvement over the first order when a global optimisable drive is applied and up to two orders of magnitude improvement when the corner spins are driven separately at short times ($\tau = 0.001J^{-1}$).  We then run the optimisations for larger systems at time $\tau = 0.1J^{-1}$ and find that this advantage is retained even with increasing system size. 

This is a big improvement over recent results in digitized adiabatic evolution with LCD \cite{sun2022optimizing}, where optimisation was used to determine optimal coefficients for second order LCD in order to prepare a GHZ state on an Ising spin chain. At 10 spins the final state fidelity for $\tau = 1J^{-1}$ obtained in their paper was 0.18, while we reach a fidelity of 0.72 for 15 spins when using corner optimisation at $\tau = 0.1J^{-1}$. 

This example shows that COLD can be used to speed up protocols which generate entanglement and is further evidence for the benefits of experimenting with different optimal control methods such as GRAPE as well as optimisation algorithms like dual annealing.

\section{Minimisation of higher order LCD terms}\label{sec:minimisation}

As alluded to in Sec.~\ref{sec:1dIsing}, the results plotted in Fig.~\ref{fig:MaxAmp} indicate that in optimising the control pulse through the parameters $\betabb$ we maximise the largest amplitude of the first order LCD drive and simultaneously reduce the second order drives.  In the Ising spin chain case this corresponds to increasing the largest amplitude of $\alpha(\lambda, \betabb)$ in Eq.~\eqref{eq:SecondLCD} throughout the evolution while reducing the maximum amplitude of both $\gamma(\lambda, \betabb)$ and $\zeta(\lambda, \betabb)$.  These results are a further indication that the implementation of COLD through the minimisation of the second-order corrections discussed in Sec.~\ref{sec:TwoSpin} may be fruitful in more complex and/or larger systems, where the dynamics can not be calculated.  

We thus investigate replacing the original cost function of Eq.~\eqref{eq:lossfunc} with one that depends a) explicitly on the maximum amplitude of the second-order drives $\gamma(\lambda, \betabb)$ and $\zeta(\lambda, \betabb)$ and b) one that depends on the total power for either drive.  Given that the LCD drives are functions of $\betabb$, one can imagine that if there is indeed a relationship between minimising a higher order drive and how effective the lower order drive is in producing the target state as a result, then we can determine parameters of the control drive that lead to a better final state fidelity.  

We take the Ising Hamiltonian from Eq.~\eqref{eq:h0_ising} and supplement it again with the parameterised control pulse from Eq.~\eqref{eq:optimsable_1}.  We once again take our first-order LCD drive to be of the form $\alpha(\lambda, \betabb) \sum_j\sigma_j^y$ and the second-order drives to be $\gamma(\lambda, \betabb) \sum_j(\sigma^x_j\sigma^y_{j+1} + \sigma^y_j\sigma^x_{j+1})$ and $\zeta(\lambda, \betabb) \sum_j(\sigma^z_j\sigma^y_{j+1} + \sigma^y_j\sigma^z_{j+1})$. In Fig.~\ref{fig:Int_Min}(a) we show the results when the cost function used to optimise the parameters $\betabb$ is the integral from Eq.~\eqref{eq:integral2} which captures the total power of the drive:
\begin{align}\label{eq:costfunc_integral}
\begin{split}
 \mathcal{C}(\betabb) &=  \int_0^\tau dt^\prime |\zeta(\lambda(t^\prime), \betabb)| \\
 &=  \mathcal{I}_2(\zeta(\lambda, \betabb)),
\end{split}
\end{align}
while in (c) we instead choose to minimise the largest amplitude of the drive reached throughout the evolution:
\begin{align}\label{eq:costfunc_maxamp}
 \mathcal{C}(\betabb) = \underset{t^{\prime} \in [0,\tau]}{\rm max_\beta} (|\zeta(\lambda(t^{\prime}), \betabb)|).
\end{align}
In both cases we plot the resulting final state fidelities for different evolution times $\tau$ and compare them to those obtained earlier in Fig.~\ref{fig:IsingUnconstrained}(a) for 5 spins.  The results are surprising in that while optimising for fidelity, as was done previously, outperforms second-order minimisation in both the integral and amplitude cases at most times, there is a stretch of driving times  around$\tau \in [0.05, 0.5]$ where second-order minimisation does better. This can be attributed to the fact that the parameter landscape for the new cost functions is completely different and allows for a more optimal value of $\betabb$ to be reached without being lost in some sub-optimal minimum during the optimisation.  

In Fig.~\ref{fig:Int_Min}(b) we plot the final state fidelities at evolution time $\tau = 0.1J^{-1}$  for up to 50 spins in order to check how this type of optimisation scales with system size and to compare the performance of both cost functions.  We find that minimising one of the two second-order drives while driving with the other still leads to impressive fidelities, but not as good as those where only first-order COLD is used. Indeed we do not have any reason to expect an absolute optimum fidelity when using this method, however, the results in Fig.~\ref{fig:Int_Min} are very encouraging.

While the new cost functions in Eqs.~\eqref{eq:costfunc_integral} and \eqref{eq:costfunc_maxamp} may seem like a roundabout way to get to the same result - a better final state fidelity in shorter time - they have several particularly important advantages over the cost function given by Eq.~\eqref{eq:lossfunc}. First and foremost, this approach does not require access to the wavefunction or experimental data at any point of the optimisation process.  In optimising for final state fidelity directly we must compute the evolution of the system many times over in order to extract the fidelity at each iteration, but computing the drive integrals or their amplitudes is completely independent of the state of the system. This allows us to determine an optimal set of parameters $\betabb$ for an arbitrary system size extremely efficiently when compared to methods which require access to $\ket{\psi_f}$. A single optimisation in their case, depending on the method used and the desired quality of the final outcome, may take hours or even days for larger system sizes.  The new method allows us to perform an optimisation with good results within minutes regardless of the number of spins, only requiring the wavefunction in order to check the resulting fidelity after the optimisation is finished.  This is a very useful tool given that most optimal control methods demand access to the wavefunction while sacrificing efficiency. 

It is not obvious that such a relationship between lower- and higher-order COLD as well as the fidelity of the final state must exist. In fact, this may be a fruitful new research direction to explore, combining the results obtained in this work along with, \@e.g.~the methods in \cite{claeys_floquet-engineering_2019}, where an approximate gauge potential can be systematically built up as a series of nested commutators. This might be a way to determine which operator ansatz $\mathcal{O}_{\rm LCD}$ has a maximal amplitude for each driven Hamiltonian and lead to a systematic optimisation of control pulses without ever having to simulate the system evolution.  There is clearly a lot of new territory to explore both in terms of optimal control and in understanding the adiabatic gauge potential a little better.

\section{Discussion and outlook} \label{sec:conclusions}

We have introduced a new hybrid approach combining quantum optimal control and shortcuts to adiabaticity: COLD. Inspired by the successes of LCD, where diabatic transitions are suppressed and locality conditions can be met, COLD improves on its methodology by combining it with quantum optimal control. The natural way to enhance the performance of LCD is by introducing higher order CD terms, but these are often non-local and difficult to engineer in experiments. COLD circumvents this by allowing for additional control fields that extend the family of dynamical Hamiltonians which can be explored. In this way, our method may find the best possible path where the effect of lower-order LCD is most relevant and higher order corrections are suppressed. 

COLD has a clear potential in efficiently speeding up adiabatic evolution in various settings. We demonstrate this numerically via several example protocols which indicate improvements beyond a classical optimisation approach BPO as well as LCD of different orders. Our work shows that COLD reduces the strength of higher order LCD corrections, and that it performs well for increasing system sizes. We have shown that COLD can be implemented in the limit of fast driving by a `switching off' of the original dynamical Hamiltonian.  For scenarios where removing the Hamiltonian is not possible, we have shown that an alternative way to implement COLD is to use a bounded optimisation where amplitudes are restricted. We find that both the COLD and COLD-CRAB protocols perform extremely well in this setting. 

COLD will be most beneficial when the LCD is only realisable to a certain order but the higher order corrections are large. This means the diabatic transitions are not being sufficiently suppressed by the choice of LCD and COLD can be used to find the dynamical Hamiltonian for which the required order of LCD term dominates. Note, that this goes the other way too, with COLD not providing substantial improvements when the chosen lower order LCD is small across the path. This can be thought of as being the case in two limits. First is the adiabatic limit, for which any CD correction is small and COLD will tend towards the adiabatic result. Second, the low-order LCD terms can be small compared to the driving as the exact CD would be correcting transitions due to interactions at longer ranges. In this scenario, the order of LCD being implemented with COLD needs to be increased, so that the CD term is accounting for the longer range terms.  We show this in Sec.~\ref{sec:ghz} where it is clear that the generation of  correlations or entanglement requires the suppression of diabatic terms that are non-local and thus first-order COLD cannot achieve a notable speed-up. In this case, higher-order corrections would need to be implemented with COLD, and finding methods for executing these non-local terms will be beneficial in these scenarios.

A further option is to combine COLD with one of a large variety of numerical optimal control methods,  as we have done for the example of CRAB and GRAPE. We have shown a substantial improvement for state preparation in the Ising model that can be obtained from the COLD-CRAB combination - particularly in the constrained case. Fusions of COLD with advanced optimal control methods for complex systems could prove even more fruitful with further study.

Another finding of our work is that COLD can be applied to more complex systems where exact dynamics are not possible, \@e.g. due to an excessively large Hilbert space. This may be achieved by variationally minimising the integrals and maximum amplitude of the driving coefficients for the higher order corrections to the LCD.  This opens up a brand new research direction as it allows for the possibility to optimise the system's path without requiring access to the system's wavefunction or any sort of experimental resource. Note, this finding is more general than COLD itself, as it can even be used to optimise protocols that do not implement LCD terms, i.e. the menagerie of control procedures currently in use, providing a cost function that does not scale with the system size. This would be implemented by minimising the highest orders of the LCD in order to find a path which allows for the least diabatic losses.

\begin{acknowledgements}
Work at the University of Strathclyde was supported by the EPSRC Quantum Technologies Hub for Quantum Computing and Simulation (EP/T001062/1), and the European Union’s Horizon 2020 research and innovation program under grant agreement No. 817482 PASQuanS.  A.P.  acknowledges support from NSF under Grant DMR-2103658 and by the AFOSR under Grants No. FA9550-16-1-0334 and FA9550-21-1-0342.

\end{acknowledgements}

\appendix

\section{Derivation of local counterdiabatic driving terms for the Ising model}\label{app:derivation}

\begin{figure*}[t]
	\includegraphics[width=0.95\linewidth]{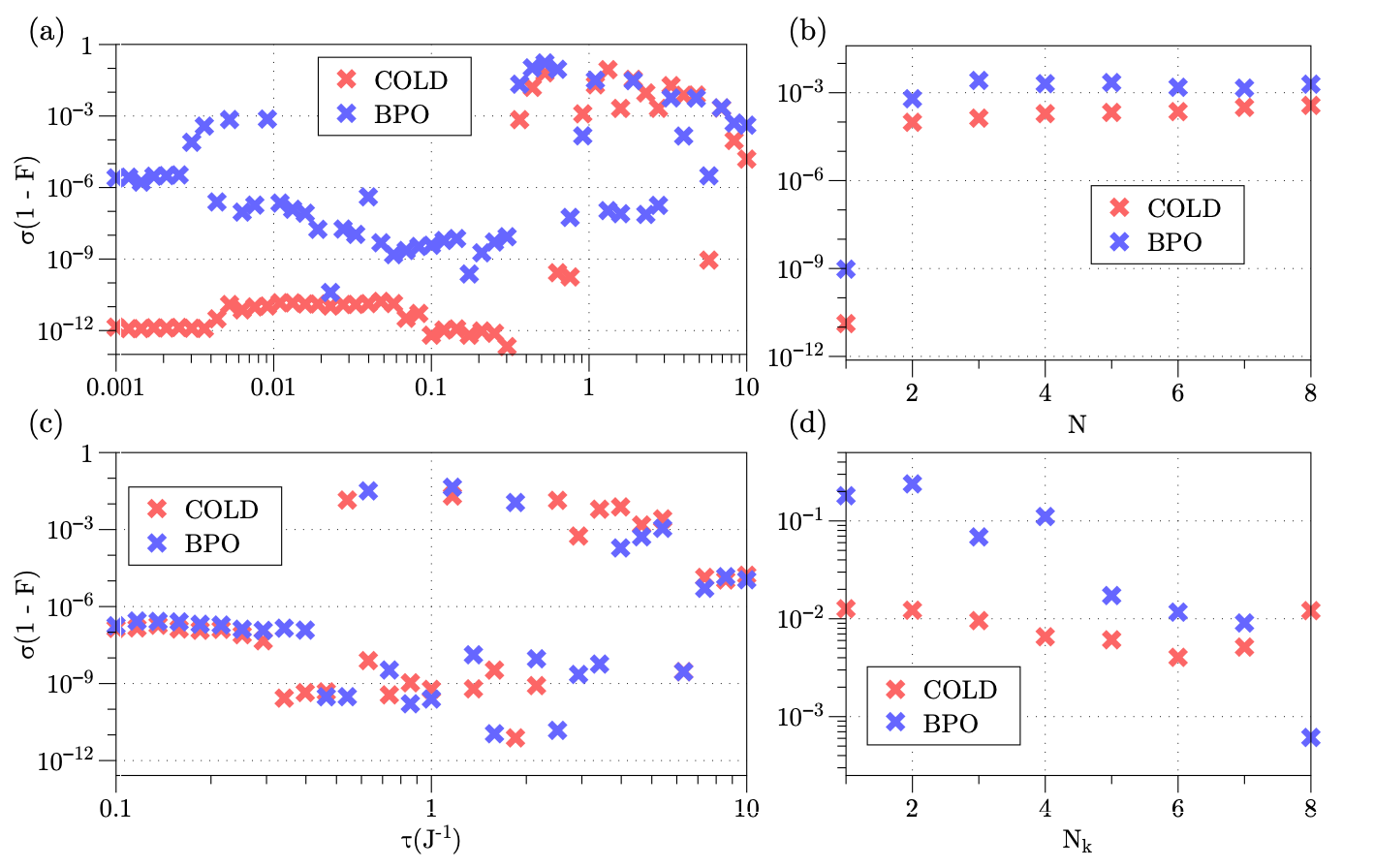}
     \caption{Plot of the standard deviations from the mean for fidelities after 500 optimisations in the case of COLD and BPO for the Ising spin chain as discussed in Sec.~\ref{sec:1dIsing}. (a) depicts the unconstrained case while (b) shows the constrained case. These correspond to the best result plots in Figs.~\ref{fig:IsingUnconstrained}(a) and \ref{fig:IsingcConstrained} (a) respectively.  The plot in (c) gives the standard deviation for increasing lengths of the chain for spin number $N$ while (d) shows the same for increasing number of parameters $N_k$. As in Fig.~\ref{fig:ScalingN}, both (b) and (c) are plotted for driving time $\tau = 10^{-2}J^{-1}$. In all plots, results for COLD are depicted with red crosses while those for BPO are depicted with blue crosses.} \label{fig:stds}
\end{figure*}

We will consider here the derivation of the coupled set of equations to be solved for the second-order LCD of the Ising model, from this, it is possible to reach all terms quoted in the main text for the examples considered.  We will consider a finite size chain of size $N$. We take the Hamiltonian to be of the general form
\begin{equation}
H = -J \sum_{j=1}^{N-1} \sigma_j^z \sigma_{j+1}^z + Z \sum_{j=1}^N \sigma_j^z + X \sum_{j=1}^N \sigma_j^x,
\end{equation}
where we will consider each coefficient to be homogeneous across the chain and dependent upon the scaling factor of $\lambda$ which is itself time-dependent as noted in the main text. We take the second order ansatz of the LCD to be that given by Eq.~\eqref{eq:SecondLCD}. We then want to obtain $G_{\lambda}$ as given by Eq.~\eqref{eq:Goperator}, which requires utilisation of standard commutation rules and the commutation relations of the Pauli matrices. Following several pages of working,  the following form of $G_{\lambda}$ can be obtained
\begin{widetext}
\begin{equation}
\begin{aligned}
G_\lambda & = - \left(\dot{J} + 4 X \zeta \right) \sum_{j=1}^{N-1} \sigma_j^z \sigma_{j+1}^z + \left(\dot{Z} + 2X\alpha \right) \sum_{j=1}^N \sigma_j^z + \left( \dot{X} - 2\alpha Z + 4 J \zeta \right) \sum_{j=1}^N \sigma_j^x + 4 J \zeta \sum_{j=1}^{N-2} \sigma_j^z \sigma_{j+1}^x \sigma_{j+2}^z \\ & + \left( 2 J \alpha + 2 X \gamma - 2 Z \zeta \right) \sum_{j=1}^{N-1} \left( \sigma_j^x \sigma_{j+1}^z + \sigma_j^z \sigma_{j+1}^x \right) +4\left( Z \gamma - X \zeta \right) \sum_{j=1}^{N-1} \sigma_j^y \sigma_{j+1}^y - 4 Z \gamma \sum_{j=1}^{N-1} \sigma_j^x \sigma_{j+1}^x \\ & + 2 J \gamma \sum_{j=1}^{N-2} \left(\sigma_j^x \sigma_{j+1}^z \sigma_{j+2}^z + \sigma_j^z \sigma_{j+1}^z \sigma_{j+2}^x + \sigma_j^z \sigma_{j+1}^y \sigma_{j+2}^y + \sigma_j^y \sigma_{j+1}^y \sigma_{j+2}^z  \right)
\end{aligned}.
\end{equation}
Note that the three spin terms would trivially go to zero for the two spin example considered in the main text. As Pauli operators are traceless, we can easily compute the Hilbert-Schmidt norm of $G_\lambda$ and we simply need to keep track of factors from the finite size of the lattice to get
\begin{figure*}[t]
	\includegraphics[width=0.95\linewidth]{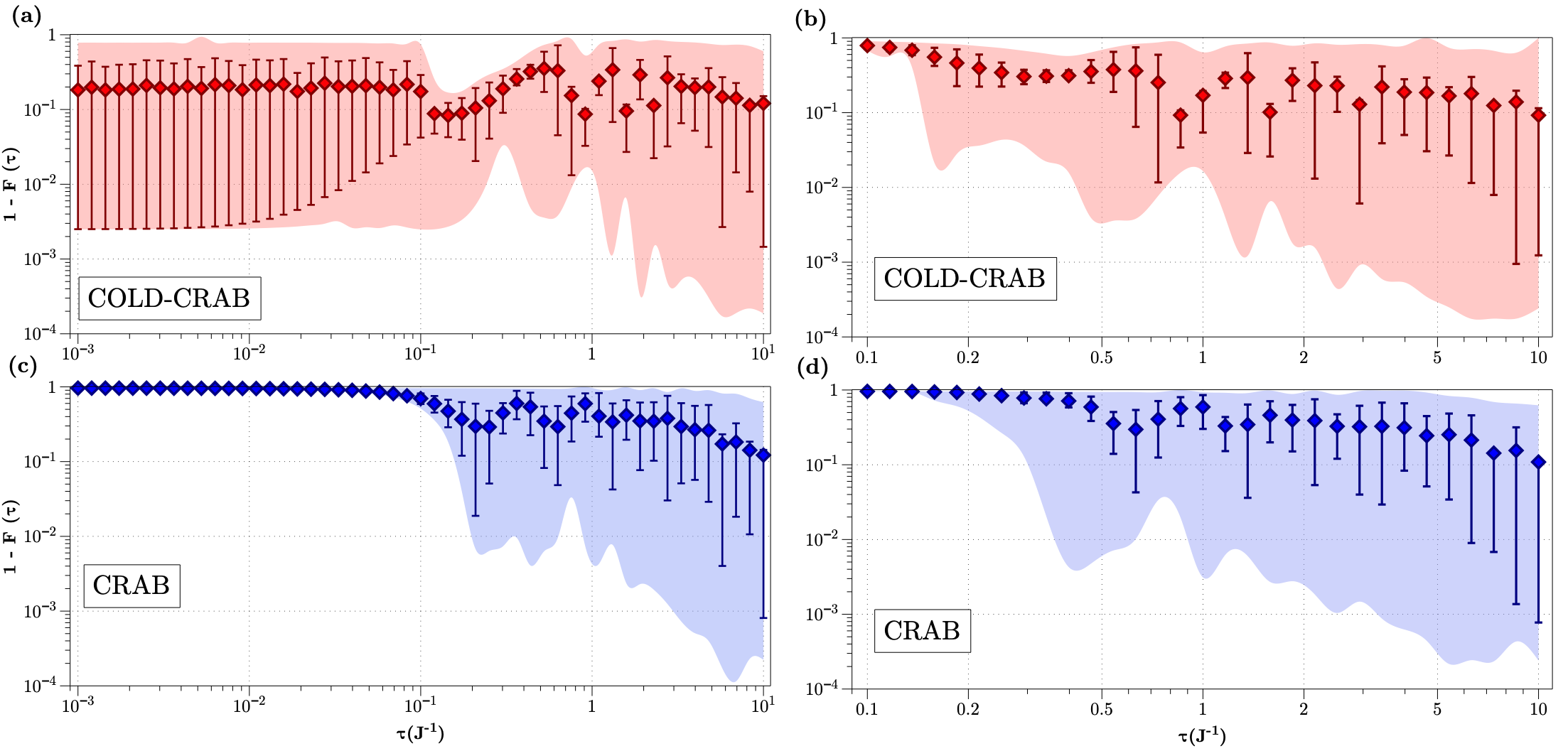}
     \caption{Plots of the mean fidelities (diamonds) obtained over 500 optimisation runs for the Ising spin chain as discussed in the main text. The error bars represent the interquartile range of the data while the shaded region encompasses the minimum and maximum fidelities obtained at each driving time.  (a) shows the case of COLD-CRAB for the constrained instance, (b) plots the same for CRAB with (c) and (d) showing results for the unconstrained Ising chain case.}\label{fig:Ising_dist}
\end{figure*}
\begin{equation}
\begin{aligned}
2^{-N}\mathrm{Tr}\left(G^2_\lambda\right) & = (N-1) \left(\dot{J} + 4 X \zeta \right)^2 + N \left(\dot{Z} + 2X\alpha \right)^2 + N \left( \dot{X} - 2\alpha Z + 4 J \zeta \right) +2 (N-1) \left( 2 J \alpha + 2 X \gamma - 2 Z \zeta \right)^2 \\ & + 16 (N-2) J^2 \gamma^2 + 16 (N-1) \left( Z \gamma - X \zeta \right)^2 + 16 (N-1) Z^2 \gamma^2 + 16 (N-2) J^2 \zeta^2
\end{aligned},
\end{equation}
where the factor on the LHS comes size of the Hilbert space. To find the system of equations to be solved we need to minimise $\mathrm{Tr}\left(G^2_\lambda\right)$ with respect to $\alpha$, $\gamma$, and $\zeta$ to obtain
\begin{equation}
\begin{aligned}
&
\begin{pmatrix}
2 \left( X^2 + Z^2 + 2\left( 1 - 1/N \right) J^2 \right) & -4 (1-1/N) & 8(1-1/N) \\ 
 -J X &  \left(  X^2 + 2(1 - \frac{1}{N-1}) J^2 + 4Z^2 \right) & -3 ZX \\
4 J Z & -6 XZ & 2 \left(4 X^2 + (4 - \frac{3}{N-1})J^2 + Z^2 \right)
\end{pmatrix}
\begin{pmatrix}
\alpha \\ \gamma \\ \zeta
\end{pmatrix}  \\ & \hspace{8cm} = 
\begin{pmatrix}
Z \dot{X} - X \dot{Z} \\ 0 \\  J \dot{X} - X \dot{J}
\end{pmatrix}
\end{aligned}.
\end{equation}
\end{widetext}
If only the first-order correction of $\alpha$ is needed, then this can be obtained by taking the first equation and setting $\gamma$ and $\zeta$ to zero. From this, the 2-spin and Ising model first order corrections can be obtained. Note in the limit of periodic boundary conditions or an infinite system we can take $N\rightarrow \infty$ to obtain the correct coefficients.  We find that the coefficients that are proportional to system size only have a significant impact when the system is very small, e.g. in the two spin case,and, therefore, have little impact the results of the Ising model with $N\geq 5$.

\section{Optimisation distributions}\label{app:distributions}

The results presented in Figs.~\ref{fig:IsingUnconstrained}, \ref{fig:IsingcConstrained} and \ref{fig:Synthetic} of the main text contain plots of the best (highest) fidelities from a number of optimisations in each instance.  Multiple optimisation runs with different initial guesses for the optimisable parameters are included to avoid pitfalls such as local minima in the parameter landscape. 

In the context of a physical implementation of one of these protocols, the optimal set of parameter values (ones which return the highest fidelity with respect to the target state) matter more than the average. However, in practice these optimisations can be very computationally costly,  in particular for larger system sizes and higher numbers of parameters.  This means that we need to understand the behaviour of the average and the worst case as they relate to the computational resources required.

If the parameter landscape is smooth and few local minima exist, then only a few optimisations are needed to determine the best values of the optimisable parameters. However, this is never a guarantee and particularly in the case of the CRAB protocol (along with COLD-CRAB), the behaviour of the optimisation is suboptimal when it comes to the number of optimisations needed to determine the parameter values which return the best fidelity of the target state.  This is due to the fact that we \emph{modify} the parameter landscape for every optimisation by randomly changing the frequency components in the control field.  While this allows each optimisation to access a new solution space and thus increases the chances of converging to a more optimal form of the control field, it also increases variance in optimisation outcomes. Since we cannot know which frequency gives the best results a priori, the only way to really reap the benefits of CRAB and CRAB-enhanced COLD is to perform as many optimisations as possible.

This can be readily seen when we look at the standard deviation in the final fidelities over all optimisations.  Fig. ~\ref{fig:stds} depicts these for the Ising spin chain of Sec.~\ref{sec:1dIsing}, both in the unconstrained and constrained case as well as for varying number of spins $N$ and parameters $N_k$.  We can see that in most cases for COLD the standard deviation of the fidelities stays below $10^{-3}$ barring longer driving times in the unconstrained case in Fig.~\ref{fig:stds}(a) as well as some in (b) for the constrained case. BPO generally displays slightly higher standard deviations, but neither shows very significant variations in the results post-optimisation.  Note that the small variation in fidelity for increasing number of parameters in Fig.~\ref{fig:stds}(d) gives further evidence for the fact that additional parameters do not improve the results of COLD or BPO in the case of the Ising chain.

When it comes to CRAB and COLD-CRAB, however, the picture is quite different.  We find that the resulting fidelities are a lot more varied across optimisations, as would be expected given the additional component of randomness. Fig.~\ref{fig:Ising_dist} shows not only the mean fidelities across optimisations but also the interquartile range of the data and the maximum and minimum values for each driving time.  We find that across optimisations we are just as likely --and in some cases far more likely -- to get a much worse final fidelity as we are to get a better one.  This is reflected in the large range between the maximum (worst) and minimum (best) fidelity for both methods as well as the interquartile range, which shows that the mean fidelity is a result of a large variation between large and small fidelities rather than a convergence to some inbetween value.

These results are useful in an assessment of computational resources for such optimisations as well as giving an insight into the range of possible outcomes, particularly when implementing more unpredicatble optimal control methods like CRAB.

\section{Choice of LCD Ansatz}\label{app:ansatz}

In determining the optimal choice of operator basis $\mathcal{O}_{\rm LCD}$, we turn back to Eq.~\eqref{eq:AGP_eq} and note that it gives us some clues about the form of the LCD. Firstly, we note that if we know nothing about the system other than, say, that it is a spin chain described by Pauli matrices, then we take the first order LCD to be all one-body terms while the second order can be two body terms and so on.  This is a natural choice due to the locality of the terms but also with respect to their practical implementation in an experiment. Given these considerations, it makes sense that for a system of spins,the first order LCD is a set of local $\sigma^y$ terms.

To illustrate, in the case of the Ising spin chain case, we know that all wave functions have real coefficients, so we know that the exact CD is given by entirely imaginary terms. We can confirm this by attempting to use local $\sigma^x$ or $\sigma^z$ terms as our ansatz for the operator basis $\mathcal{O}_{\rm LCD}$ and find that their coeffcients are equal to $0$ throughout the driving time. In the case of ansatz $\mathcal{O}_{\rm LCD} = \alpha_x \sum_j \sigma^x_j$ we find:
\begin{equation}
  \begin{split}
    G_{\lambda, \alpha_x} &= \dot{J} \sum_{j}^{N-1} \sigma^z_j\sigma^z_{j+1} + \dot{X} \sum_j^N \sigma^x_j + \dot{Z} \sum_j^N \sigma^z_j \\
    &+ 2\alpha_x J \sum_{j}^{N-1} \sigma^y_j\sigma^z_{j+1} + 2\alpha_x J \sum_{j}^{N-1} \sigma^z_j\sigma^y_{j+1} \\ 
    &+ 2\alpha_x Z \sum_j \sigma^y_j,
    \end{split}
\end{equation}
according to Eq.~\eqref{eq:Goperator}, Then the action, as in Eq.~\eqref{eq:actionCD}, is:
\begin{equation}
    \begin{split}
        \mathcal{S}(\mathcal{A}_{\lambda}) &= 2^{-N}\Trace{\left[G_{\lambda, \alpha_x}(\mathcal{A}_{\lambda})^2\right]} \\
        &= (1 - \frac{1}{N})\dot{J}^2 + \dot{X}^2 + \dot{Z}^2 + (1 - \frac{1}{N})8\alpha_x^2 J^2 \\ 
        &+ 4\alpha_x^2 Z^2,
    \end{split}
\end{equation}
which, when minimised with respect to $\alpha_x$ gives $\alpha_x = 0$. 

The same procedure can be done for $\mathcal{O}_{\rm LCD} = \alpha_z \sum_j \sigma^z_j$:
\begin{equation}
  \begin{split}
    G_{\lambda, \alpha_z} &= \dot{J} \sum_{j}^{N-1} \sigma^z_j\sigma^z_{j+1} + \dot{X} \sum_j^N \sigma^x_j + \dot{Z} \sum_j^N \sigma^z_j \\
    &- 2 \alpha_z X \sum_j^N \sigma_j^y,
    \end{split}
\end{equation}
where again we take the action:
\begin{equation}
    \begin{split}
        \mathcal{S}(\mathcal{A}_{\lambda}) &= 2^{-N}\Trace{\left[G_{\lambda, \alpha_x}(\mathcal{A}_{\lambda})^2\right]} \\
        &= (1 - \frac{1}{N})\dot{J}^2 + \dot{X}^2 + \dot{Z}^2 + 4\alpha_z^2 X^2,
    \end{split}
\end{equation}
which minimised with respect to $\alpha_z$ once again gives $\alpha_z = 0$.

\bibliography{mybib}

\end{document}